\documentclass[prd,aps,nofootinbib,floatfix,11pt]{revtex4}

\usepackage{amsmath,graphicx,epsfig,amssymb,dsfont,mathtools}
\usepackage[usenames]{color}
\usepackage{ulem} 
\usepackage{bigstrut}
\usepackage{slashed}
\usepackage{multirow}
\usepackage{subfigure}

\allowdisplaybreaks

\begin{document}
	
	\title{Weak Decays of Heavy Baryons in Light-Front Approach}
	
	\author{Zhen-Xing Zhao$^{1}$~\thanks{Email:star\_0027@sjtu.edu.cn}}
	
	\affiliation{$^{1}$ INPAC, Shanghai Key Laboratory for Particle Physics and Cosmology,
		\\
		MOE Key Laboratory for Particle Physics, Astrophysics and Cosmology,
		\\
		School of Physics and Astronomy, Shanghai Jiao-Tong University, Shanghai
		200240, P.R. China }
		
	\begin{abstract}
		In this work, we perform a analysis of semi-leptonic and nonleptonic weak decays of
		  heavy baryons: $\Lambda_{b},\Xi_{b},\Omega_{b}$  and $\Lambda_{c},\Xi_{c},\Omega_{c}$. For nonleptonic decay modes, we study only the factorizable channels induced by the external W-emission.
		  The two spectator quarks in baryonic transitions are treated as a
		diquark and form factors are calculated in the light-front approach.
	Using the results for form factors, we also calculate some corresponding  semi-leptonic and nonleptonic decay widths. We find that our results are
		comparable with the available experimental data and other theoretical
		predictions. Decay branching fractions for many channels are found
		to reach the level $10^{-3}\sim10^{-2}$, which are promising to be
		discovered in the future measurements at BESIII, LHCb and BelleII.
		The SU(3) symmetry in semi-leptonic decays is examined and sources
		of symmetry breaking are discussed. 
	\end{abstract}
	\maketitle
	
	\section{Introduction}
	
	Quite recently, the LHCb collaboration announced the discovery of
	the doubly charmed baryon $\Xi_{cc}^{++}$~\cite{Aaij:2017ueg}.
	Undoubtedly this discovery will open a new door to study strong interactions
	in the presence of a pair of heavy quarks. Accordingly it has triggered
	great theoretical interests in studying doubly heavy baryons from
	different aspects~\cite{Chen:2017sbg,Yu:2017zst,Wang:2017mqp,Li:2017cfz,Meng:2017udf,Wang:2017azm,Karliner:2017qjm,Gutsche:2017hux,Li:2017pxa,Guo:2017vcf,Lu:2017meb,Xiao:2017udy,Sharma:2017txj,Ma:2017nik,Meng:2017dni,Li:2017ndo,Wang:2017qvg,Shi:2017dto,Hu:2017dzi,1803.01476}.
	
	Inspired by this discovery, we also expect a renaissance in the study
	of singly bottom or charm baryons. Particularly there are rapid progresses
	in the study of $\Lambda_{c}$ decays at BESIII~\cite{Ablikim:2015flg,Ablikim:2015prg,Ablikim:2016mcr,Ablikim:2016vqd,Ablikim:2017iqd,Ablikim:2017ors}
	and some recent studies on $\Lambda_{b}$ and $\Lambda_{c}$ decays
	by LHCb can be found in Refs.~\cite{Aaij:2017awb,Aaij:2017nsd,Aaij:2017pgy,Aaij:2017rin,Aaij:2017svr,Aaij:2017xva}. It is anticipated that 
	many more decay modes will be established in future. Thus an up-to-date
	theoretical analysis is highly demanded, and this work aims to do
	so.
	
\begin{figure}
		\includegraphics[width=0.5\columnwidth]{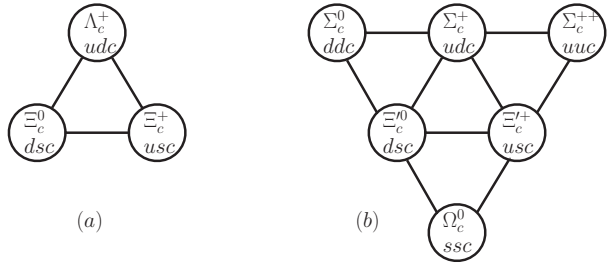} \caption{Anti-triplets (panel a) and sextets (panel b) of charmed baryons with
			one charm quark and two light quarks. It is similar for the baryons
			with a bottom quark. }
		\label{fig:singly_heavy} 
	\end{figure}

	Quark model is a very successful tool in classifying mesons and baryons.
	A heavy baryon is composed of one heavy quark $c/b$ and two light
	quarks. Light flavor SU(3) symmetry arranges the singly heavy baryons
	into the presentations $\boldsymbol{3}\otimes\boldsymbol{3}=\boldsymbol{6}\oplus\bar{\boldsymbol{3}}$,
	as can be seen from Fig.~\ref{fig:singly_heavy}. For charmed baryons,
	the irreducible representation $\bar{\boldsymbol{3}}$ is composed
	of $\Lambda_{c}^{+}$ and $\Xi_{c}^{+,0}$ while the sextet is composed
	of $\Sigma_{c}^{++,+,0}$, $\Xi_{c}^{\prime+,\prime0}$ and $\Omega_{c}^{0}$.
	They all have spin $1/2$, but only 4 of them weakly decay predominantly:
	$\Lambda_{c}^{+}$ and $\Xi_{c}^{+,0}$ in the representation $\bar{\boldsymbol{3}}$
	and $\Omega_{c}^{0}$ in the representation $\boldsymbol{6}$. Others
	can decays into the lowest-lying states via strong or electromagnetic
	interactions. This is similar for bottomed baryons. In this work,
	we will focus on weak decays of singly heavy baryons and more explicitly
	we will consider only the following channels: 
	\begin{itemize}
		\item charm sector: 
		\begin{eqnarray*}
			\Lambda_{c}^{+}(cud) & \to & n(dud)/\Lambda(sud),\\
			\Xi_{c}^{+}(cus) & \to & \Sigma^{0}(dus)/\Lambda(dus)/\Xi^{0}(sus),\\
			\Xi_{c}^{0}(cds) & \to & \Sigma^{-}(dds)/\Xi^{-}(sds),\\
			\Omega_{c}^{0}(css) & \to & \Xi^{-}(dss);
		\end{eqnarray*}
		\item bottom sector: 
		\begin{eqnarray*}
			\Lambda_{b}^{0}(bud) & \to & p(uud)/\Lambda_{c}^{+}(cud),\\
			\Xi_{b}^{0}(bus) & \to & \Sigma^{+}(uus)/\Xi_{c}^{+}(cus),\\
			\Xi_{b}^{-}(bds) & \to & \Sigma^{0}(uds)/\Lambda(uds)/\Xi_{c}^{0}(cds),\\
			\Omega_{b}^{-}(bss) & \to & \Xi^{0}(uss)/\Omega_{c}^{0}(css).
		\end{eqnarray*}
	\end{itemize}
	In the above, we have listed the quark contents of the baryons in
	the brakets and placed the quarks that participate in weak decay in
	the first place.
	
	The light baryons in the final state are composed of 3 light quarks
	and belong to the baryon octet. Their wave functions, including the
	flavor and spin spaces, have the form~\cite{Halzen:1984mc} 
	\begin{equation}
	{\cal B}_{\boldsymbol{8}}=\sqrt{\frac{1}{2}}({\rm p}_{S}\chi(M_{S})+{\rm p}_{A}\chi(M_{A})).
	\end{equation}
	Here ${\rm p}_{S(A)}$ stands for the mixed symmetric (antisymmetric)
	$\boldsymbol{8}$ in the SU(3) representation decmposition $\boldsymbol{3}\otimes\boldsymbol{3}\otimes\boldsymbol{3}=\boldsymbol{10}\oplus\boldsymbol{8}\oplus\boldsymbol{8}\oplus\boldsymbol{1}$
	in the flavor space, while $\chi(M_{S(A)})$ stands for the mixed
	symmetric (antisymmetric) $\boldsymbol{2}$ in the SU(2) representation
	decmposition $\boldsymbol{2}\otimes\boldsymbol{2}\otimes\boldsymbol{2}=\boldsymbol{4}\oplus\boldsymbol{2}\oplus\boldsymbol{2}$
	in the spin space. Here the ``mixed symmetric (antisymmetric)\char`\"{}
	means the state is symmetric (antisymmetric) under interchange of
	the first two quarks. The wave functions for baryons in the initial
	and final states are collected in the Appendix~\ref{app:wave_functions}.
	
\begin{figure}[!]
		\includegraphics[width=0.4\columnwidth]{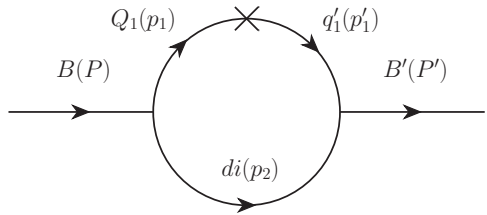} \caption{Feynman diagrams for baryon-baryon transitions in the diquark picture.
			$P^{(\prime)}$ is the momentum of the incoming (outgoing) baryon,
			$p_{1}^{(\prime)}$ is the initial (final) quark momentum, $p_{2}$
			is the diquark momentum and the cross mark denotes the corresponding
			vertex of weak interaction.}
		\label{fig:decay} 
	\end{figure}

	On the theoretical side, the singly heavy baryon decays have been
	investigated by various theoretical methods, and some of them can
	be found in Refs.~\cite{Cheng:1991sn,Gronau:2013mza,He:2015fwa,Savage:1989qr,Faustov:2016yza,Cheng:2018hwl,Li:2016qai,Cheng:2015ckx,Zhu:2016bra,Guo:2005qa,He:2006ud,Wang:2008sm,Lu:2009cm,Wang:2009hra,Khodjamirian:2011jp,Wang:2015ndk,Detmold:2015aaa,Meinel:2016dqj,Meinel:2017ggx,Huber:2016xod,Feldmann:2011xf,Boer:2014kda}.
	In this work, we will adopt the light-front approach. This method
	has been widely used to study the properties of mesons~\cite{Jaus:1999zv,Jaus:1989au,Jaus:1991cy,Cheng:1996if,Cheng:2003sm,Cheng:2004yj,Ke:2009ed,Ke:2009mn,Cheng:2009ms,Lu:2007sg,Wang:2007sxa,Wang:2008xt,Wang:2008ci,Wang:2009mi,Chen:2009qk,Li:2010bb,Verma:2011yw,Shi:2016gqt}.
	Its application to baryons can be found in Refs.~\cite{Ke:2007tg,Wei:2009np,Ke:2012wa,Zhu:2018jet}.
	In the transition form factors, the two spectator quarks do not change
	and can be viewed as a diquark. In this diquark scheme, the two quarks
	are treated as a whole system, and thus its role is similar to that
	of the antiquark in the meson case, see Fig.~\ref{fig:decay}. In
	the process like $\Lambda_{b}\to\Lambda_{c}$, where the light quarks
	u and d are considered to form a scalar diquark, which is denoted
	by {[}ud{]}, while in the process like $\Omega_{b}\to\Omega_{c}$,
	the light s quarks are believed to form an axial-vector diquark, which
	is denoted by \{ss\}.
	
	Some recent works have been devoted to investigate the singly heavy
	baryon decays with the help of flavor SU(3) symmetry~\cite{Lu:2016ogy,Wang:2017gxe,Geng:2017mxn,Geng:2018plk}.
	Based on the available data, the SU(3) analysis can give predictions
	on a great number of decay modes ranging from semi-leptonic decays
	to multi-body nonleptonic decays. However, as we know, in the case
	of c quark decay, SU(3) symmetry breaking effects are sizable and
	can not be omitted. A quantitative study of SU(3) symmetry breaking
	effects will be conducted within the light-front approach.
	
	The rest of the paper is arranged as follows. In Sec. II, we will
	present briefly the framework of light-front approach under the diquark
	picture, and the wave function overlapping factors are also given.
	Our results are shown in Sec. III, including the results for form
	factors, predictions on semi-leptonic and nonleptonic decay widths,
	and detailed discussions on the SU(3) symmetry and sources of symmetry
	breaking. A brief summary will be given in the last section.
	
	\section{Theoretical framework}
	
	In this section, we will briefly overview the theoretical framework
	for form factors: the light-front approach. More details can
	be found in Refs.~\cite{Ke:2007tg} and \cite{Wang:2017mqp}. It
	is necessary to point out that the physical form factor should be
	multiplied by a factor due to the overlap of wave funtions in the
	initial and final states.
	
	\subsection{Form factors}
	
	\label{subsec:ff}
	
	The transition matrix elements are parameterized as 
	\begin{eqnarray}
	\langle B^{\prime}(P^{\prime},S_{z}^{\prime})|V_{\mu}|B(P,S_{z})\rangle & = & \bar{u}(P^{\prime},S_{z}^{\prime})\left[\gamma_{\mu}f_{1}(q^{2})+i\sigma_{\mu\nu}\frac{q^{\nu}}{M}f_{2}(q^{2})+\frac{q_{\mu}}{M}f_{3}(q^{2})\right]u(P,S_{z}),\nonumber \\
	\langle B^{\prime}(P^{\prime},S_{z}^{\prime})|A_{\mu}|B(P,S_{z})\rangle & = & \bar{u}(P^{\prime},S_{z}^{\prime})\left[\gamma_{\mu}g_{1}(q^{2})+i\sigma_{\mu\nu}\frac{q^{\nu}}{M}g_{2}(q^{2})+\frac{q_{\mu}}{M}g_{3}(q^{2})\right]\gamma_{5}u(P,S_{z}),\label{eq:weakMatrix1}
	\end{eqnarray}
	where $q=P-P^{\prime}$, $M$ denotes the mass of the parent baryon
	$B$, and $f_{i}$, $g_{i}$ are form factors.
	
	In the light-front approach, hadron states are expanded in terms of
	quark states superposed with a wave function, where the momentum and other quantum numbers are considered
	simultaneously. Then the weak transition matrix element can be obtained
	as 
	\begin{align}
	\langle B^{\prime}(P^{\prime},S_{z}^{\prime})|(V-A)_{\mu}|B(P,S_{z})\rangle & =\int\{d^{3}p_{2}\}\frac{\phi^{\prime*}(x^{\prime},k_{\perp}^{\prime})\phi(x,k_{\perp})}{2\sqrt{p_{1}^{+}p_{1}^{\prime+}(p_{1}\cdot\bar{P}+m_{1}M_{0})(p_{1}^{\prime}\cdot\bar{P}^{\prime}+m_{1}^{\prime}M_{0}^{\prime})}}\nonumber \\
	& \quad\times\bar{u}(\bar{P}^{\prime},S_{z}^{\prime})\bar{\Gamma}^{\prime}(\slashed p_{1}^{\prime}+m_{1}^{\prime})\gamma_{\mu}(1-\gamma_{5})(\slashed p_{1}+m_{1})\Gamma u(\bar{P},S_{z}).\label{eq:weakMatrix2}
	\end{align}
	
	From Eqs. (\ref{eq:weakMatrix1}) and (\ref{eq:weakMatrix2}), we
	can extract the explicit expressions of form factors: 
	\begin{align}
	f_{1}(q^{2})= & \int\frac{dxd^{2}k_{\perp}}{2(2\pi)^{3}}\frac{\phi^{\prime}(x^{\prime},k_{\perp}^{\prime})\phi(x,k_{\perp})[k_{\perp}\cdot k_{\perp}^{\prime}+(x_{1}M_{0}+m_{1})(x_{1}^{\prime}M_{0}^{\prime}+m_{1}^{\prime})]}{\sqrt{\left[(m_{1}+x_{1}M_{0})^{2}+k_{\perp}^{2}\right]\left[(m_{1}^{\prime}+x_{1}^{\prime}M_{0}^{\prime})^{2}+k_{\perp}^{\prime2}\right]}},\nonumber \\
	g_{1}(q^{2})= & \int\frac{dxd^{2}k_{\perp}}{2(2\pi)^{3}}\frac{\phi^{\prime}(x^{\prime},k_{\perp}^{\prime})\phi(x,k_{\perp})[-k_{\perp}\cdot k_{\perp}^{\prime}+(x_{1}M_{0}+m_{1})(x_{1}^{\prime}M_{0}^{\prime}+m_{1}^{\prime})]}{\sqrt{\left[(m_{1}+x_{1}M_{0})^{2}+k_{\perp}^{2}\right]\left[(m_{1}^{\prime}+x_{1}^{\prime}M_{0}^{\prime})^{2}+k_{\perp}^{\prime2}\right]}},\nonumber \\
	\frac{f_{2}(q^{2})}{M}= & \frac{1}{q_{\perp}^{2}}\int\frac{dxd^{2}k_{\perp}}{2(2\pi)^{3}}\frac{\phi^{\prime}(x^{\prime},k_{\perp}^{\prime})\phi(x,k_{\perp})[-(m_{1}+x_{1}M_{0})k_{\perp}^{\prime}\cdot q_{\perp}+(m_{1}^{\prime}+x_{1}^{\prime}M_{0}^{\prime})k_{\perp}\cdot q_{\perp}]}{\sqrt{\left[(m_{1}+x_{1}M_{0})^{2}+k_{\perp}^{2}\right]\left[(m_{1}^{\prime}+x_{1}^{\prime}M_{0}^{\prime})^{2}+k_{\perp}^{\prime2}\right]}},\nonumber \\
	\frac{g_{2}(q^{2})}{M}= & \frac{1}{q_{\perp}^{2}}\int\frac{dxd^{2}k_{\perp}}{2(2\pi)^{3}}\frac{\phi^{\prime}(x^{\prime},k_{\perp}^{\prime})\phi(x,k_{\perp})[-(m_{1}+x_{1}M_{0})k_{\perp}^{\prime}\cdot q_{\perp}-(m_{1}^{\prime}+x_{1}^{\prime}M_{0}^{\prime})k_{\perp}\cdot q_{\perp}]}{\sqrt{\left[(m_{1}+x_{1}M_{0})^{2}+k_{\perp}^{2}\right]\left[(m_{1}^{\prime}+x_{1}^{\prime}M_{0}^{\prime})^{2}+k_{\perp}^{\prime2}\right]}}\label{eq:ff_scalar}
	\end{align}
	for a scalar diquark involved in the initial and final baryons, or
	\begin{align}
	f_{1}(q^{2})= & \frac{1}{8P^{+}P^{\prime+}}\int\frac{dx_{2}d^{2}k_{\perp}}{2(2\pi)^{3}}\frac{\varphi^{\prime}(x^{\prime},k_{\perp}^{\prime})\varphi(x,k_{\perp})}{6\sqrt{x_{1}x_{1}^{\prime}(p_{1}\cdot\bar{P}+m_{1}M_{0})(p_{1}^{\prime}\cdot\bar{P}^{\prime}+m_{1}^{\prime}M_{0}^{\prime})}}\nonumber \\
	& \times{\rm Tr}[(\bar{\slashed P}+M_{0})\gamma^{+}(\bar{\slashed P^{\prime}}+M_{0}^{\prime})\gamma_{5}\gamma_{\alpha}(\slashed p_{1}^{\prime}+m_{1}^{\prime})\gamma^{+}(\slashed p_{1}+m_{1})\gamma_{5}\gamma_{\beta}](\frac{p_{2}^{\alpha}p_{2}^{\beta}}{m_{2}^{2}}-g^{\alpha\beta}),\nonumber \\
	g_{1}(q^{2})= & \frac{1}{8P^{+}P^{\prime+}}\int\frac{dx_{2}d^{2}k_{\perp}}{2(2\pi)^{3}}\frac{\varphi^{\prime}(x^{\prime},k_{\perp}^{\prime})\varphi(x,k_{\perp})}{2\sqrt{x_{1}x_{1}^{\prime}(p_{1}\cdot\bar{P}+m_{1}M_{0})(p_{1}^{\prime}\cdot\bar{P}^{\prime}+m_{1}^{\prime}M_{0}^{\prime})}}\nonumber \\
	& \times{\rm Tr}[(\bar{\slashed P}+M_{0})\gamma^{+}\gamma_{5}(\bar{\slashed P^{\prime}}+M_{0}^{\prime})\gamma_{5}\gamma_{\alpha}(\slashed p_{1}^{\prime}+m_{1}^{\prime})\gamma^{+}\gamma_{5}(\slashed p_{1}+m_{1})\gamma_{5}\gamma_{\beta}](\frac{p_{2}^{\alpha}p_{2}^{\beta}}{m_{2}^{2}}-g^{\alpha\beta}),\nonumber \\
	\frac{f_{2}(q^{2})}{M}= & -\frac{1}{8P^{+}P^{\prime+}}\frac{iq_{\perp}^{i}}{q_{\perp}^{2}}\int\frac{dx_{2}d^{2}k_{\perp}}{2(2\pi)^{3}}\frac{\varphi^{\prime}(x^{\prime},k_{\perp}^{\prime})\varphi(x,k_{\perp})}{2\sqrt{x_{1}x_{1}^{\prime}(p_{1}\cdot\bar{P}+m_{1}M_{0})(p_{1}^{\prime}\cdot\bar{P}^{\prime}+m_{1}^{\prime}M_{0}^{\prime})}}\nonumber \\
	& \times{\rm Tr}[(\bar{\slashed P}+M_{0})\sigma^{i+}(\bar{\slashed P^{\prime}}+M_{0}^{\prime})\gamma_{5}\gamma_{\alpha}(\slashed p_{1}^{\prime}+m_{1}^{\prime})\gamma^{+}(\slashed p_{1}+m_{1})\gamma_{5}\gamma_{\beta}](\frac{p_{2}^{\alpha}p_{2}^{\beta}}{m_{2}^{2}}-g^{\alpha\beta}),\nonumber \\
	\frac{g_{2}(q^{2})}{M}= & \frac{1}{8P^{+}P^{\prime+}}\frac{iq_{\perp}^{i}}{q_{\perp}^{2}}\int\frac{dx_{2}d^{2}k_{\perp}}{2(2\pi)^{3}}\frac{\varphi^{\prime}(x^{\prime},k_{\perp}^{\prime})\varphi(x,k_{\perp})}{2\sqrt{x_{1}x_{1}^{\prime}(p_{1}\cdot\bar{P}+m_{1}M_{0})(p_{1}^{\prime}\cdot\bar{P}^{\prime}+m_{1}^{\prime}M_{0}^{\prime})}}\nonumber \\
	& \times{\rm Tr}[(\bar{\slashed P}+M_{0})\sigma^{i+}\gamma_{5}(\bar{\slashed P^{\prime}}+M_{0}^{\prime})\gamma_{5}\gamma_{\alpha}(\slashed p_{1}^{\prime}+m_{1}^{\prime})\gamma^{+}\gamma_{5}(\slashed p_{1}+m_{1})\gamma_{5}\gamma_{\beta}](\frac{p_{2}^{\alpha}p_{2}^{\beta}}{m_{2}^{2}}-g^{\alpha\beta})\label{eq:ff_axial}
	\end{align}
	if an axial-vector diquark is involved.
	
	\subsection{Spin and flavor wave functions}
	
	In the last subsection, we have presented the explicit expressions
	of form factors. It should be noted that, the physical transition
	form factor should be multiplied by the corresponding overlapping
	factor: 
	\begin{equation}
	f_{1}^{{\rm physical}}(q^{2})=\text{the overlapping factor}\times f_{1}^{\text{in Sec.~\ref{subsec:ff}}}(q^{2}).\label{eq:physical_overlapping}
	\end{equation}
	From the discussions in the Appendix \ref{app:wave_functions}, we
	can obtain these factors and the corresponding results are collected
	in Tab.~\ref{Tab:overlappings}.
	
	\begin{table}[!htb]
		\caption{The overlapping factors in the transitions}
		\label{Tab:overlappings} %
		\begin{tabular}{c|c}
			\hline 
			transitions  & overlapping factors\tabularnewline
			\hline 
			$\Lambda_{c}^{+}(cud)\to n(dud)/\Lambda(sud)$  & $-\frac{1}{\sqrt{2}},\quad\frac{1}{\sqrt{3}}$\tabularnewline
			\hline 
			$\Xi_{c}^{+}(cus)\to\Sigma^{0}(dus)/\Lambda(dus)/\Xi^{0}(sus)$  & $\frac{1}{2},\quad\frac{1}{2\sqrt{3}},\quad-\frac{1}{\sqrt{2}}$\tabularnewline
			\hline 
			$\Xi_{c}^{0}(cds)\to\Sigma^{-}(dds)/\Xi^{-}(sds)$  & $\frac{1}{\sqrt{2}},\quad-\frac{1}{\sqrt{2}}$\tabularnewline
			\hline 
			$\Omega_{c}^{0}(css)\to\Xi^{-}(dss)$  & $-\frac{1}{\sqrt{3}}$\tabularnewline
			\hline 
			$\Lambda_{b}^{0}(bud)\to p(uud)/\Lambda_{c}^{+}(cud)$  & $\frac{1}{\sqrt{2}},\quad1$\tabularnewline
			\hline 
			$\Xi_{b}^{0}(bus)\to\Sigma^{+}(uus)/\Xi_{c}^{+}(cus)$  & $\frac{1}{\sqrt{2}},\quad1$\tabularnewline
			\hline 
			$\Xi_{b}^{-}(bds)\to\Sigma^{0}(uds)/\Lambda(uds)/\Xi_{c}^{0}(cds)$  & $\frac{1}{2},\quad-\frac{1}{2\sqrt{3}},\quad1$\tabularnewline
			\hline 
			$\Omega_{b}^{-}(bss)\to\Xi^{0}(uss)/\Omega_{c}^{0}(css)$  & $-\frac{1}{\sqrt{3}},\quad1$\tabularnewline
			\hline 
		\end{tabular}
	\end{table}

	\section{Numerical results and discussions}
	
	All   inputs to calculate the form factors will be collected in the first subsection. What follows is the numerical results for form factors, semi-leptonic and
	nonleptonic processes. Some remarks will also be given.
	The last subsection is devoted to an SU(3) analysis for semi-leptonic
	processes.
	
	\subsection{Inputs}
	
	The quark masses used in the model are given as 
	\[
	m_{u}=m_{d}=0.25{\rm GeV},\quad m_{s}=0.37{\rm GeV},\quad m_{c}=1.4{\rm GeV},\quad m_{b}=4.8{\rm GeV}.
	\]
	These values are widely adopted in Refs.~\cite{Lu:2007sg,Wang:2007sxa,Wang:2008xt,Wang:2008ci,Wang:2009mi,Chen:2009qk,Li:2010bb,Verma:2011yw,Shi:2016gqt}.
	The diquark masses are chosen as 
	\begin{eqnarray*}
		&  & m_{[ud]}=0.50{\rm GeV},\quad m_{[us]}=m_{[ds]}=0.60{\rm GeV},\\
		&  & m_{\{uu\}}=m_{\{ud\}}=m_{\{dd\}}=0.77{\rm GeV},\quad m_{\{us\}}=m_{\{ds\}}=0.87{\rm GeV},\quad m_{\{ss\}}=0.97{\rm GeV}.
	\end{eqnarray*}
	Here the square brackets (curly braces) denote a scalar (an axial-vector)
	diquark.
	
	The shape parameters are given as 
	\begin{eqnarray*}
		&  & \beta_{b[ud]}=0.66{\rm GeV},\quad\beta_{b[us]}=\beta_{b[ds]}=0.68{\rm GeV},\quad\beta_{b\{ss\}}=0.78{\rm GeV},\\
		&  & \beta_{c[ud]}=0.56{\rm GeV},\quad\beta_{c[us]}=\beta_{c[ds]}=0.58{\rm GeV},\quad\beta_{c\{ss\}}=0.66{\rm GeV},\\
		&  & \beta_{s[ud]}=0.45{\rm GeV},\quad\beta_{s[us]}=\beta_{s[ds]}=0.46{\rm GeV},\\
		&  & \beta_{d[ud]}=0.40{\rm GeV},\quad\beta_{d[us]}=\beta_{d[ds]}=0.41{\rm GeV},\quad\beta_{d\{ss\}}=0.44{\rm GeV},\\
		&  & \beta_{u[ud]}=0.40{\rm GeV},\quad\beta_{u[us]}=\beta_{u[ds]}=0.41{\rm GeV},\quad\beta_{u\{ss\}}=0.44{\rm GeV}.
	\end{eqnarray*}
	
	Some remarks on the above parameters are given in order. 
	\begin{itemize}
		\item $m_{[ud]}=0.50{\rm GeV}$ and $m_{\{ud\}}=0.77{\rm GeV}$ are taken
		from Refs.~\cite{Ke:2007tg,Ke:2012wa}. Other diquark masses are taken
		as the above values since the s quark mass is expected to be 0.1 GeV
		higher than that of u or d quark. 
		\item The shape parameters for baryons are constrained by the corresponding
		ones for mesons~\cite{Cheng:2003sm}. To be specific, $\beta_{b,di}$
		are taken as between $\beta_{b\bar{s}}=0.623$ and $\beta_{b\bar{c}}=0.886$;
		$\beta_{c,di}$ are taken as between $\beta_{c\bar{s}}=0.535$ and
		$\beta_{c\bar{c}}=0.753$; $\beta_{d,di}$ are taken as between $\beta_{d\bar{s}}=0.393$
		and $\beta_{d\bar{c}}=0.470$; $\beta_{s,di}$ are taken as between
		$\beta_{s\bar{s}}=0.440$ and $\beta_{s\bar{c}}=0.535$. 
	\end{itemize}
	The masses and lifetimes of the parent baryons are collected in Tab.~\ref{Tab:parent_baryons}
	and the masses of the daughter baryons are shown in Tab.~\ref{Tab:daughter_baryons}~\cite{Olive:2016xmw}.
	
	\begin{table}[!htb]
		\caption{Masses and lifetimes of parent baryons. $\Lambda$ and $\Xi$ are
			in the representation $\bar{\boldsymbol{3}}$ while $\Omega$ are
			in the representation $\boldsymbol{6}$. }
		\label{Tab:parent_baryons} %
		\begin{tabular}{c|c|c|c|c||c|c|c|c}
			\hline 
			baryon  & $\Lambda_{c}^{+}$  & $\Xi_{c}^{+}$  & $\Xi_{c}^{0}$  & $\Omega_{c}^{0}$  & $\Lambda_{b}^{0}$  & $\Xi_{b}^{0}$  & $\Xi_{b}^{-}$  & $\Omega_{b}^{-}$\tabularnewline
			\hline 
			mass/GeV  & 2.286  & 2.468  & 2.471  & 2.695  & 5.620  & 5.792  & 5.795  & 6.046\tabularnewline
			\hline 
			lifetime/fs  & 200  & 442  & 112  & 69  & 1466  & 1464  & 1560  & 1570\tabularnewline
			\hline 
		\end{tabular}
	\end{table}
	
	\begin{table}[!htb]
		\caption{Masses of daughter baryons. They form the baryon octet. }
		\label{Tab:daughter_baryons} %
		\begin{tabular}{c|c|c|c|c|c|c|c|c}
			\hline 
			baryon  & $p$  & $n$  & $\Lambda$  & $\Sigma^{+}$  & $\Sigma^{0}$  & $\Sigma^{-}$  & $\Xi^{0}$  & $\Xi^{-}$\tabularnewline
			\hline 
			mass/GeV  & 0.938  & 0.940  & 1.116  & 1.189  & 1.193  & 1.197  & 1.315  & 1.322\tabularnewline
			\hline 
		\end{tabular}
	\end{table}
	
	Fermi constant and CKM matrix elements are also taken from PDG~\cite{Olive:2016xmw}:
	\begin{align}
	& G_{F}=1.166\times10^{-5}{\rm GeV}^{-2},\nonumber \\
	& |V_{ud}|=0.974,\quad|V_{us}|=0.225,\quad|V_{ub}|=0.00357,\nonumber \\
	& |V_{cd}|=0.225,\quad|V_{cs}|=0.974,\quad|V_{cb}|=0.0411.\label{eq:GFCKM}
	\end{align}

	\subsection{Form factors}
	
	Results for form factors are collected in Tab.~\ref{Tab:ff_charm}
	for charmed baryons and Tab.~\ref{Tab:ff_bottom} for bottomed baryons.
	The following expressions have been used to access the $q^2$ distribution:  
	\begin{equation}
	F(q^{2})=\frac{F(0)}{1\mp\frac{q^{2}}{m_{{\rm fit}}^{2}}+\delta\left(\frac{q^{2}}{m_{{\rm fit}}^{2}}\right)^{2}},\label{eq:fit_formula}
	\end{equation}
	where the $F(0)$ is the form factor at $q^{2}=0$. The $m_{{\rm fit}}$
	and $\delta$ are two parameters to be fitted from numerical results.
	For the form factor $g_{2}$, a plus sign is adopted in Eq.~(\ref{eq:fit_formula})
	otherwise the fitted parameter $m_{{\rm fit}}$ becomes purely imaginary.
	The minus sign is adopted for all the other situations.
	
	Some comments are given in order. 
	\begin{itemize}

\item Only the scalar diquark contributes to the $\Lambda_{Q}$ and $\Xi_{Q}$ decays and only the axial-vector diquark contributes to the $\Omega_{Q}$ decays, where $Q=c/b$.
		\item It should be noted that, in Tabs.~\ref{Tab:ff_charm} and \ref{Tab:ff_bottom},
		the overlapping factors are not taken into account. The physical transition
		form factor should be multiplied by the corresponding overlapping factor, see
		Eq.~(\ref{eq:physical_overlapping}). 
		\item An advantage of the results in Tab.~\ref{Tab:ff_charm} is that,
		they can be directly be used to explore SU(3) symmetry and its breaking
		effects. In fact, if we take the approximations 
		\begin{eqnarray*}
			&  & m_{d}=m_{s},\\
			&  & m_{[ud]}=m_{[us]}=m_{[ds]}=m_{\{ss\}},\\
			&  & \beta_{c[ud]}=\beta_{c[us]}=\beta_{c[ds]}=\beta_{c\{ss\}},\\
			&  & \beta_{d[ud]}=\beta_{s[ud]}=\beta_{d[us]}=\beta_{s[us]}=\beta_{d[ds]}=\beta_{s[ds]}=\beta_{d\{ss\}}
		\end{eqnarray*}
		and 
		\[
		m_{\Lambda_{c}^{+}}=m_{\Xi_{c}^{+}}=m_{\Xi_{c}^{0}}=m_{\Omega_{c}^{0}},
		\]
		all the form factors will be the same. From the results in Tab.~\ref{Tab:ff_charm}, we can see
		that the SU(3) symmetry is not severely broken.
	\end{itemize}

In Tab. \ref{Tab:comparison_ff}, we compare our results with
other theoretical predictions in Refs. \cite{Faustov:2016yza,Gutsche:2014zna,Liu:2009sn}. Some comments are
given as follows.
\begin{itemize}
\item In Tab. \ref{Tab:comparison_ff}, the physcial form factors are shown, see Eq. (\ref{eq:physical_overlapping}).
\item It can be seen that, our results are comparable to other predictions.
However, there still exists an uncertainty about the sign of $g_{2}(0)$.
The sign of $g_{2}(0)$ in this work is same as that derived by LCSR
method in Ref. \cite{Liu:2009sn} but different from those obtained by other quark
models. More careful analysis should be devoted to fixing this problem.
\item The form factors $f_{3}$ and $g_{3}$ are not obtained in this work
because we have taken the $q^{+}=0$ frame, while another method adopted
in Refs. \cite{Ke:2017eqo,Zhao:2018mrg} may be applied to extract these form factors.
\item Also note that, in Refs. \cite{Faustov:2016yza,Gutsche:2014zna,Liu:2009sn}, only few channels are investigated
but this work aims to give a comprehensive investigation to the heavy
baryon decays. Only in this way, SU(3) symmetry and sources of SU(3)
symmetry breaking can be seen clearly.
\end{itemize}

\begin{table}
\caption{A comparison with other results for the form factors at the maximum
recoil $q^{2}=0$. The physical form factors are shown in ``this
work'' with the help of Eq. (\ref{eq:physical_overlapping}).}
\label{Tab:comparison_ff}

\begin{tabular}{lcccccc}
\hline 
 & $f_{1}(0)$  & $f_{2}(0)$  & $f_{3}(0)$ & $g_{1}(0)$  & $g_{2}(0)$  & $g_{3}(0)$\tabularnewline
\hline 
$\Lambda_{c}\to n$  &  &  &  &  &  & \tabularnewline
this work & $0.513$  & $-0.266$  & - - & $0.443$ & $-0.034$ & - -\tabularnewline
Quark model \cite{Faustov:2016yza} & $0.627$  & $-0.259$  & $0.179$ & $0.433$  & $0.118$  & $-0.744$\tabularnewline
Quark model \cite{Gutsche:2014zna} & $0.470$  & $-0.246$  & $0.039$ & $0.414$  & $0.073$  & $-0.328$ \tabularnewline
\hline 
$\Lambda_{c}\to\Lambda$  &  &  &  &  &  & \tabularnewline
this work  & $0.468$  & $-0.222$  & - - & $0.407$  & $-0.035$  & - -\tabularnewline
Quark model \cite{Faustov:2016yza} & $0.700$  & $-0.295$  & $0.222$ & $0.448$  & $0.135$  & $-0.832$\tabularnewline
Quark model \cite{Gutsche:2014zna} & $0.511$  & $-0.289$  & $-0.014$ & $0.466$  & $0.025$  & $-0.400$\tabularnewline
LCSR \cite{Liu:2009sn} & $0.517$  & $-0.123$  & - - & $0.517$  & $-0.123$  & - -\tabularnewline
\hline
\end{tabular}
\end{table}

	\begin{table}
		\caption{Form factors for charmed baryon decays. The plus sign is adopted in
			the fit formula Eq.~(\ref{eq:fit_formula}) for the ones with asterisk,
			and the minus sign is adopted for all the others.}
		\label{Tab:ff_charm} %
		\begin{tabular}{c|c|c|c|c|c|c|c}
			\hline 
			$F$  & $F(0)$  & $m_{{\rm fit}}$  & $\delta$  & $F$  & $F(0)$  & $m_{{\rm fit}}$  & $\delta$\tabularnewline
			\hline 
			$f_{1}^{\Lambda_{c}^{+}\to n}$  & $0.726$  & $2.06$  & $0.07$  & $f_{2}^{\Lambda_{c}^{+}\to n}$  & $-0.376$  & $1.71$  & $0.21$ \tabularnewline
			$g_{1}^{\Lambda_{c}^{+}\to n}$  & $0.626$  & $2.40$  & $0.19$  & $g_{2}^{\Lambda_{c}^{+}\to n}$  & $-0.048$  & $1.42$  & $0.29$ \tabularnewline
			\hline 
			$f_{1}^{\Lambda_{c}^{+}\to\Lambda}$  & $0.810$  & $2.19$  & $0.06$  & $f_{2}^{\Lambda_{c}^{+}\to\Lambda}$  & $-0.384$  & $1.81$  & $0.20$ \tabularnewline
			$g_{1}^{\Lambda_{c}^{+}\to\Lambda}$  & $0.705$  & $2.52$  & $0.16$  & $g_{2}^{\Lambda_{c}^{+}\to\Lambda}$  & $-0.060$  & $1.58$  & $0.25$ \tabularnewline
			\hline 
			$f_{1}^{\Xi_{c}^{+}\to\Sigma^{0}}$  & $0.717$  & $2.00$  & $0.11$  & $f_{2}^{\Xi_{c}^{+}\to\Sigma^{0}}$  & $-0.421$  & $1.69$  & $0.22$ \tabularnewline
			$g_{1}^{\Xi_{c}^{+}\to\Sigma^{0}}$  & $0.614$  & $2.34$  & $0.22$  & $g_{2}^{\Xi_{c}^{+}\to\Sigma^{0}}$  & $-0.054$  & $1.39$  & $0.32$ \tabularnewline
			\hline 
			$f_{1}^{\Xi_{c}^{+}\to\Lambda}$  & $0.717$  & $2.00$  & $0.11$  & $f_{2}^{\Xi_{c}^{+}\to\Lambda}$  & $-0.421$  & $1.69$  & $0.22$ \tabularnewline
			$g_{1}^{\Xi_{c}^{+}\to\Lambda}$  & $0.614$  & $2.34$  & $0.22$  & $g_{2}^{\Xi_{c}^{+}\to\Lambda}$  & $-0.054$  & $1.39$  & $0.32$ \tabularnewline
			\hline 
			$f_{1}^{\Xi_{c}^{+}\to\Xi^{0}}$  & $0.802$  & $2.13$  & $0.10$  & $f_{2}^{\Xi_{c}^{+}\to\Xi^{0}}$  & $-0.431$  & $1.79$  & $0.22$ \tabularnewline
			$g_{1}^{\Xi_{c}^{+}\to\Xi^{0}}$  & $0.694$  & $2.46$  & $0.19$  & $g_{2}^{\Xi_{c}^{+}\to\Xi^{0}}$  & $-0.065$  & $1.55$  & $0.28$ \tabularnewline
			\hline 
			$f_{1}^{\Xi_{c}^{0}\to\Sigma^{-}}$  & $0.717$  & $2.00$  & $0.11$  & $f_{2}^{\Xi_{c}^{0}\to\Sigma^{-}}$  & $-0.422$  & $1.69$  & $0.22$ \tabularnewline
			$g_{1}^{\Xi_{c}^{0}\to\Sigma^{-}}$  & $0.614$  & $2.34$  & $0.22$  & $g_{2}^{\Xi_{c}^{0}\to\Sigma^{-}}$  & $-0.054$  & $1.39$  & $0.32$ \tabularnewline
			\hline 
			$f_{1}^{\Xi_{c}^{0}\to\Xi^{-}}$  & $0.802$  & $2.13$  & $0.10$  & $f_{2}^{\Xi_{c}^{0}\to\Xi^{-}}$  & $-0.432$  & $1.79$  & $0.22$ \tabularnewline
			$g_{1}^{\Xi_{c}^{0}\to\Xi^{-}}$  & $0.694$  & $2.46$  & $0.19$  & $g_{2}^{\Xi_{c}^{0}\to\Xi^{-}}$  & $-0.065$  & $1.55$  & $0.28$ \tabularnewline
			\hline 
			$f_{1}^{\Omega_{c}^{0}\to\Xi^{-}}$  & $0.653$  & $1.51$  & $0.35$  & $f_{2}^{\Omega_{c}^{0}\to\Xi^{-}}$  & $0.620$  & $1.54$  & $0.28$ \tabularnewline
			$g_{1}^{\Omega_{c}^{0}\to\Xi^{-}}$  & $-0.182$  & $1.92$  & $0.08$  & $g_{2}^{\Omega_{c}^{0}\to\Xi^{-}}$  & $0.002^{*}$  & $0.40^{*}$  & $0.14^{*}$ \tabularnewline
			\hline 
		\end{tabular}
	\end{table}
	
	\begin{table}
		\caption{Same as Tab.~\ref{Tab:ff_charm} but for the bottomed baryon case. }
		\label{Tab:ff_bottom} %
		\begin{tabular}{c|c|c|c|c|c|c|c}
			\hline 
			$F$  & $F(0)$  & $m_{{\rm fit}}$  & $\delta$  & $F$  & $F(0)$  & $m_{{\rm fit}}$  & $\delta$\tabularnewline
			\hline 
			$f_{1}^{\Lambda_{b}^{0}\to p}$  & $0.282$  & $4.66$  & $0.30$  & $f_{2}^{\Lambda_{b}^{0}\to p}$  & $-0.084$  & $3.94$  & $0.37$ \tabularnewline
			$g_{1}^{\Lambda_{b}^{0}\to p}$  & $0.273$  & $4.81$  & $0.32$  & $g_{2}^{\Lambda_{b}^{0}\to p}$  & $-0.012$  & $3.67$  & $0.37$ \tabularnewline
			\hline 
			$f_{1}^{\Lambda_{b}^{0}\to\Lambda_{c}^{+}}$  & $0.670$  & $5.62$  & $0.23$  & $f_{2}^{\Lambda_{b}^{0}\to\Lambda_{c}^{+}}$  & $-0.132$  & $4.67$  & $0.32$ \tabularnewline
			$g_{1}^{\Lambda_{b}^{0}\to\Lambda_{c}^{+}}$  & $0.656$  & $5.73$  & $0.24$  & $g_{2}^{\Lambda_{b}^{0}\to\Lambda_{c}^{+}}$  & $-0.012$  & $4.37$  & $0.28$ \tabularnewline
			\hline 
			$f_{1}^{\Xi_{b}^{0}\to\Sigma^{+}}$  & $0.260$  & $4.46$  & $0.34$  & $f_{2}^{\Xi_{b}^{0}\to\Sigma^{+}}$  & $-0.086$  & $3.84$  & $0.40$ \tabularnewline
			$g_{1}^{\Xi_{b}^{0}\to\Sigma^{+}}$  & $0.251$  & $4.60$  & $0.36$  & $g_{2}^{\Xi_{b}^{0}\to\Sigma^{+}}$  & $-0.012$  & $3.56$  & $0.41$ \tabularnewline
			\hline 
			$f_{1}^{\Xi_{b}^{0}\to\Xi_{c}^{+}}$  & $0.654$  & $5.42$  & $0.27$  & $f_{2}^{\Xi_{b}^{0}\to\Xi_{c}^{+}}$  & $-0.143$  & $4.59$  & $0.34$ \tabularnewline
			$g_{1}^{\Xi_{b}^{0}\to\Xi_{c}^{+}}$  & $0.640$  & $5.53$  & $0.28$  & $g_{2}^{\Xi_{b}^{0}\to\Xi_{c}^{+}}$  & $-0.015$  & $4.33$  & $0.30$ \tabularnewline
			\hline 
			$f_{1}^{\Xi_{b}^{-}\to\Sigma^{0}}$  & $0.260$  & $4.46$  & $0.34$  & $f_{2}^{\Xi_{b}^{-}\to\Sigma^{0}}$  & $-0.086$  & $3.84$  & $0.40$ \tabularnewline
			$g_{1}^{\Xi_{b}^{-}\to\Sigma^{0}}$  & $0.251$  & $4.60$  & $0.36$  & $g_{2}^{\Xi_{b}^{-}\to\Sigma^{0}}$  & $-0.012$  & $3.56$  & $0.41$ \tabularnewline
			\hline 
			$f_{1}^{\Xi_{b}^{-}\to\Lambda}$  & $0.260$  & $4.46$  & $0.34$  & $f_{2}^{\Xi_{b}^{-}\to\Lambda}$  & $-0.086$  & $3.84$  & $0.40$ \tabularnewline
			$g_{1}^{\Xi_{b}^{-}\to\Lambda}$  & $0.251$  & $4.60$  & $0.36$  & $g_{2}^{\Xi_{b}^{-}\to\Lambda}$  & $-0.012$  & $3.56$  & $0.41$ \tabularnewline
			\hline 
			$f_{1}^{\Xi_{b}^{-}\to\Xi_{c}^{0}}$  & $0.654$  & $5.42$  & $0.27$  & $f_{2}^{\Xi_{b}^{-}\to\Xi_{c}^{0}}$  & $-0.143$  & $4.59$  & $0.34$ \tabularnewline
			$g_{1}^{\Xi_{b}^{-}\to\Xi_{c}^{0}}$  & $0.640$  & $5.53$  & $0.28$  & $g_{2}^{\Xi_{b}^{-}\to\Xi_{c}^{0}}$  & $-0.015$  & $4.33$  & $0.30$ \tabularnewline
			\hline 
			$f_{1}^{\Omega_{b}^{-}\to\Xi^{0}}$  & $0.169$  & $3.30$  & $0.64$  & $f_{2}^{\Omega_{b}^{-}\to\Xi^{0}}$  & $0.193$  & $3.45$  & $0.49$ \tabularnewline
			$g_{1}^{\Omega_{b}^{-}\to\Xi^{0}}$  & $-0.033$  & $4.38$  & $0.20$  & $g_{2}^{\Omega_{b}^{-}\to\Xi^{0}}$  & $-0.041$  & $4.32$  & $0.65$ \tabularnewline
			\hline 
			$f_{1}^{\Omega_{b}^{-}\to\Omega_{c}^{0}}$  & $0.566$  & $3.92$  & $0.49$  & $f_{2}^{\Omega_{b}^{-}\to\Omega_{c}^{0}}$  & $0.531$  & $4.08$  & $0.41$ \tabularnewline
			$g_{1}^{\Omega_{b}^{-}\to\Omega_{c}^{0}}$  & $-0.170$  & $4.80$  & $0.23$  & $g_{2}^{\Omega_{b}^{-}\to\Omega_{c}^{0}}$  & $-0.031$  & $9.02$  & $5.05$ \tabularnewline
			\hline 
		\end{tabular}
	\end{table}

	\subsection{Semi-leptonic results}
	
	The differential decay width for semi-leptonic process reads 
	\begin{equation}
	\frac{d\Gamma}{dq^{2}}=\frac{d\Gamma_{L}}{dq^{2}}+\frac{d\Gamma_{T}}{dq^{2}},
	\end{equation}
	with the polarized decay widths are given as 
	\begin{align}
	\frac{d\Gamma_{L}}{dq^{2}} & =\frac{G_{F}^{2}|V_{CKM}|^{2}}{(2\pi)^{3}}\frac{q^{2}p}{24M^{2}}(|H_{\frac{1}{2},0}|^{2}+|H_{-\frac{1}{2},0}|^{2}),\\
	\frac{d\Gamma_{T}}{dq^{2}} & =\frac{G_{F}^{2}|V_{CKM}|^{2}}{(2\pi)^{3}}\frac{q^{2}p}{24M^{2}}(|H_{\frac{1}{2},1}|^{2}+|H_{-\frac{1}{2},-1}|^{2}).
	\end{align}
	Here the $q^{2}$ is the lepton pair invariant mass, $p=\sqrt{Q_{+}Q_{-}}/2M$, $Q_{\pm}=(M\pm M^{\prime})^{2}-q^{2}$, and $M$ ($M^{\prime}$) is the mass of the parent (daughter) baryon.
	
	The helicity amplitudes are related to the form factors through the
	following expressions: 
	\begin{align}
	H_{\frac{1}{2},0}^{V} & =-i\frac{\sqrt{Q_{-}}}{\sqrt{q^{2}}}\left((M+M^{\prime})f_{1}-\frac{q^{2}}{M}f_{2}\right),\nonumber \\
	H_{\frac{1}{2},1}^{V} & =i\sqrt{2Q_{-}}\left(-f_{1}+\frac{M+M^{\prime}}{M}f_{2}\right),\nonumber \\
	H_{\frac{1}{2},0}^{A} & =-i\frac{\sqrt{Q_{+}}}{\sqrt{q^{2}}}\left((M-M^{\prime})g_{1}+\frac{q^{2}}{M}g_{2}\right),\nonumber \\
	H_{\frac{1}{2},1}^{A} & =i\sqrt{2Q_{+}}\left(-g_{1}-\frac{M-M^{\prime}}{M}g_{2}\right).
	\end{align}
	The negative helicity amplitudes are
	given as 
	\begin{equation}
	H_{-\lambda^{\prime},-\lambda_{W}}^{V}=H_{\lambda^{\prime},\lambda_{W}}^{V}\quad\text{and}\quad H_{-\lambda^{\prime},-\lambda_{W}}^{A}=-H_{\lambda^{\prime},\lambda_{W}}^{A}.
	\end{equation}
	The helicity amplitudes for the left-handed current are obtained as
	\begin{equation}
	H_{\lambda^{\prime},\lambda_{W}}=H_{\lambda^{\prime},\lambda_{W}}^{V}-H_{\lambda^{\prime},\lambda_{W}}^{A}.
	\end{equation}
	
	Numerical results are given in Tabs.~\ref{Tab:semi_charm} and
	\ref{Tab:semi_bottom}. Comparisons with some recent works \cite{Geng:2018plk,Geng:2017mxn,Gutsche:2014zna,Meinel:2017ggx,Gutsche:2015mxa,Meinel:2016dqj,Detmold:2015aaa} and the
	experimental results \cite{Olive:2016xmw} can be found in Tabs.~\ref{Tab:comparison_semi_charm}
	and \ref{Tab:comparison_semi_bottom}.
	
	\begin{table}
		\caption{Semi-leptonic decays for charmed baryons. }
		\label{Tab:semi_charm} %
		\begin{tabular}{l|c|c|c}
			\hline 
			channels  & $\Gamma/{\rm GeV}$  & ${\cal B}$  & $\Gamma_{L}/\Gamma_{T}$\tabularnewline
			\hline 
			$\Lambda_{c}^{+}\to ne^{+}\nu_{e}$  & $6.62\times10^{-15}$  & $2.01\times10^{-3}$  & $1.78$\tabularnewline
			$\Lambda_{c}^{+}\to\Lambda e^{+}\nu_{e}$  & $5.36\times10^{-14}$  & $1.63\times10^{-2}$  & $1.96$\tabularnewline
			$\Xi_{c}^{+}\to\Sigma^{0}e^{+}\nu_{e}$  & $2.79\times10^{-15}$  & $1.87\times10^{-3}$  & $1.85$\tabularnewline
			$\Xi_{c}^{+}\to\Lambda e^{+}\nu_{e}$  & $1.22\times10^{-15}$  & $8.22\times10^{-4}$  & $1.79$\tabularnewline
			$\Xi_{c}^{+}\to\Xi^{0}e^{+}\nu_{e}$  & $8.03\times10^{-14}$  & $5.39\times10^{-2}$  & $1.98$\tabularnewline
			$\Xi_{c}^{0}\to\Sigma^{-}e^{+}\nu_{e}$  & $5.57\times10^{-15}$  & $9.47\times10^{-4}$  & $1.86$\tabularnewline
			$\Xi_{c}^{0}\to\Xi^{-}e^{+}\nu_{e}$  & $7.91\times10^{-14}$  & $1.35\times10^{-2}$  & $1.98$\tabularnewline
			$\Omega_{c}^{0}\to\Xi^{-}e^{+}\nu_{e}$  & $2.08\times10^{-15}$  & $2.18\times10^{-4}$  & $7.94$\tabularnewline
			\hline 
		\end{tabular}
	\end{table}

	\begin{table}
		\caption{A comparison with some recent works for semi-leptonic charmed decays. }
		\label{Tab:comparison_semi_charm} %
		\begin{tabular}{c|c|c|c}
			\hline 
			channel  & this work  & other theoretical predictions & experiment~\cite{Olive:2016xmw}\tabularnewline
			\hline 
			$\Lambda_{c}^{+}\to ne^{+}\nu_{e}$  & $2.01\times10^{-3}$  & $(2.7\pm0.3)\times10^{-3}$ \cite{Geng:2017mxn}, $2.07\times10^{-3}$
			\cite{Gutsche:2014zna}, $(4.10\pm0.26)\times10^{-3}$ \cite{Meinel:2017ggx} & - -\tabularnewline
			$\Lambda_{c}^{+}\to\Lambda e^{+}\nu_{e}$  & $1.63\times10^{-2}$  & $2.72\times10^{-2}$ \cite{Gutsche:2015mxa}, $(3.80\pm0.22)\times10^{-2}$\cite{Meinel:2016dqj} & $(3.6\pm0.4)\times10^{-2}$\tabularnewline
			$\Xi_{c}^{+}\to\Sigma^{0}e^{+}\nu_{e}$  & $1.87\times10^{-3}$  & $(0.8\pm0.1)\times10^{-3}$ \cite{Geng:2017mxn}  & - -\tabularnewline
			$\Xi_{c}^{+}\to\Lambda e^{+}\nu_{e}$  & $8.22\times10^{-4}$  & $(2.5\pm0.4)\times10^{-4}$ \cite{Geng:2017mxn}  & - -\tabularnewline
			$\Xi_{c}^{+}\to\Xi^{0}e^{+}\nu_{e}$  & $5.39\times10^{-2}$  & $(3.38_{-2.26}^{+2.19})\times10^{-2}$ \cite{Geng:2018plk}, $(3.0\pm0.5)\times10^{-2}$
			\cite{Geng:2017mxn}  & - -\tabularnewline
			$\Xi_{c}^{0}\to\Sigma^{-}e^{+}\nu_{e}$  & $9.47\times10^{-4}$  & $(60\pm8)\times10^{-4}$ \cite{Geng:2017mxn}  & - -\tabularnewline
			$\Xi_{c}^{0}\to\Xi^{-}e^{+}\nu_{e}$  & $1.35\times10^{-2}$  & $(4.87\pm1.74)\times10^{-2}$ \cite{Geng:2018plk}, $(11.9\pm1.6)\times10^{-2}$
			\cite{Geng:2017mxn}  & - -\tabularnewline
			\hline 
		\end{tabular}
	\end{table}
	
	\begin{table}
		\caption{Semi-leptonic decays for bottomed baryons. }
		\label{Tab:semi_bottom} %
		\begin{tabular}{l|c|c|c}
			\hline 
			channels  & $\Gamma/{\rm GeV}$  & ${\cal B}$  & $\Gamma_{L}/\Gamma_{T}$\tabularnewline
			\hline 
			$\Lambda_{b}^{0}\to pe^{-}\bar{\nu}_{e}$  & $1.41\times10^{-16}$  & $3.14\times10^{-4}$  & $1.25$\tabularnewline
			$\Lambda_{b}^{0}\to\Lambda_{c}^{+}e^{-}\bar{\nu}_{e}$  & $3.96\times10^{-14}$  & $8.83\times10^{-2}$  & $1.71$\tabularnewline
			$\Xi_{b}^{0}\to\Sigma^{+}e^{-}\bar{\nu}_{e}$  & $1.27\times10^{-16}$  & $2.83\times10^{-4}$  & $1.27$\tabularnewline
			$\Xi_{b}^{0}\to\Xi_{c}^{+}e^{-}\bar{\nu}_{e}$  & $3.97\times10^{-14}$  & $8.83\times10^{-2}$  & $1.70$\tabularnewline
			$\Xi_{b}^{-}\to\Sigma^{0}e^{-}\bar{\nu}_{e}$  & $6.37\times10^{-17}$  & $1.51\times10^{-4}$  & $1.27$\tabularnewline
			$\Xi_{b}^{-}\to\Lambda e^{-}\bar{\nu}_{e}$  & $2.29\times10^{-17}$  & $5.42\times10^{-5}$  & $1.25$\tabularnewline
			$\Xi_{b}^{-}\to\Xi_{c}^{0}e^{-}\bar{\nu}_{e}$  & $3.97\times10^{-14}$  & $9.42\times10^{-2}$  & $1.70$\tabularnewline
			$\Omega_{b}^{-}\to\Xi^{0}e^{-}\bar{\nu}_{e}$  & $1.18\times10^{-17}$  & $2.82\times10^{-5}$  & $1.72$\tabularnewline
			$\Omega_{b}^{-}\to\Omega_{c}^{0}e^{-}\bar{\nu}_{e}$  & $1.14\times10^{-14}$  & $2.72\times10^{-2}$  & $6.26$\tabularnewline
			\hline 
		\end{tabular}
	\end{table}
	
	\begin{table}
		\caption{A comparison with some recent works for semi-leptonic bottomed decays. }
		\label{Tab:comparison_semi_bottom} %
		\begin{tabular}{c|c|c|c}
			\hline 
			channel  & this work  & other theoretical predictions & experiment~\cite{Olive:2016xmw}\tabularnewline
			\hline 
			$\Lambda_{b}^{0}\to pe^{-}\bar{\nu}_{e}$  & $3.14\times10^{-4}$  & $2.9\times10^{-4}$\cite{Gutsche:2014zna}, $(4.80\pm0.99)\times10^{-4}$\cite{Detmold:2015aaa} & $(4.1\pm1.0)\times10^{-4}$\tabularnewline
			$\Lambda_{b}^{0}\to\Lambda_{c}^{+}e^{-}\bar{\nu}_{e}$  & $8.83\times10^{-2}$  & $6.9\times10^{-2}$ \cite{Gutsche:2015mxa}, $(5.32\pm0.35)\times10^{-2}$\cite{Detmold:2015aaa} & $(6.2_{-1.3}^{+1.4})\times10^{-2}$\tabularnewline
			\hline 
		\end{tabular}
	\end{table}

	\subsection{Non-leptonic results}
	
	For nonleptonic decays, we are constrained to consider only the processes of a W boson emitting outward. The naive factorization assumption is employed \cite{Bauer:1984zv,Bauer:1986bm}. The decay width for the $B\to B^{\prime}P$ ($P$ denotes a pseudoscalar
	meson) is given as 
	\begin{equation}
	\Gamma=\frac{p}{8\pi}\left(\frac{(M+M^{\prime})^{2}-m^{2}}{M^{2}}|A|^{2}+\frac{(M-M^{\prime})^{2}-m^{2}}{M^{2}}|B|^{2}\right),
	\end{equation}
	where $p$ is the magnitude of the three-momentum of the daughter
	baryon $B^{\prime}$ in the rest frame of the parent baryon $B$.
	$M$ ($M^{\prime}$) is the mass of the parent (daughter) baryon.
	For $B\to B^{\prime}V(A)$ ($V$ denotes a vector meson while $A$
	denotes an axial-vector meson) decay, the decay width is 
	\begin{equation}
	\Gamma=\frac{p(E^{\prime}+M^{\prime})}{4\pi M}\left(2(|S|^{2}+|P_{2}|^{2})+\frac{E^{2}}{m^{2}}(|S+D|^{2}+|P_{1}|^{2})\right),
	\end{equation}
	where $E$ ($E'$) is the energy of the meson (daughter baryon) in
	the final state, and 
	\begin{align*}
	S & =-A_{1},\\
	P_{1} & =-\frac{p}{E}\left(\frac{M+M^{\prime}}{E^{\prime}+M^{\prime}}B_{1}+B_{2}\right),\\
	P_{2} & =\frac{p}{E^{\prime}+M^{\prime}}B_{1},\\
	D & =-\frac{p^{2}}{E(E^{\prime}+M^{\prime})}(A_{1}-A_{2}).
	\end{align*}
	$A$, $B$, $A_{1,2}$ and $B_{1,2}$ are given as 
	\begin{align}
	A & =-\lambda f_{P}(M-M^{\prime})f_{1}(m^{2}),\nonumber \\
	B & =-\lambda f_{P}(M+M^{\prime})g_{1}(m^{2}),\nonumber \\
	A_{1} & =-\lambda f_{V}m\left[g_{1}(m^{2})+g_{2}(m^{2})\frac{M-M^{\prime}}{M}\right],\nonumber \\
	A_{2} & =-2\lambda f_{V}mg_{2}(m^{2}),\nonumber \\
	B_{1} & =\lambda f_{V}m\left[f_{1}(m^{2})-f_{2}(m^{2})\frac{M+M^{\prime}}{M}\right],\nonumber \\
	B_{2} & =2\lambda f_{V}mf_{2}(m^{2}).\label{eq:AB}
	\end{align}
	Here $\lambda=\frac{G_{F}}{\sqrt{2}}V_{CKM}V_{q_{1}q_{2}}^{*}a_{1}$
	with $a_{1}=C_{1}(\mu_{c})+C_{2}(\mu_{c})/3=1.07$ \cite{Li:2012cfa},
	the first CKM matrix element corresponds to the process of $B\to B^{\prime}$
	and the second comes from the production of the meson. $M(M')$ is
	the mass of the parent (daughter) baryon and $m$ is the mass of the
	emitted meson. For the decay mode with an axial-vector meson involved,
	$f_{V}$ should be replaced by $-f_{A}$ in the expressions of $A_{1,2}$
	and $B_{1,2}$ in Eqs.~\eqref{eq:AB}.

Note that the P-wave meson $a_{1}$ emission case is included. The naive factorization can still work for these processes \cite{Wang:2008hu}.

	The masses of the mesons in the final states can be taken from Ref.~\cite{Olive:2016xmw}.
	The decay constants are adopted as~\cite{Cheng:2003sm,Shi:2016gqt,Carrasco:2014poa}
	\begin{align}
	f_{\pi} & =130.4{\rm MeV},\quad f_{\rho}=216{\rm MeV},\quad f_{a_{1}}=238{\rm MeV},\quad f_{K}=160{\rm MeV},\quad f_{K^{*}}=210{\rm MeV},\nonumber \\
	f_{D} & =207.4{\rm MeV},\quad f_{D^{*}}=220{\rm MeV},\quad f_{D_{s}}=247.2{\rm MeV},\quad f_{D_{s}^{*}}=247.2{\rm MeV}.
	\end{align}
	
	The numerical results are given in Tabs.~\ref{Tab:nonlep_charm},
	\ref{Tab:nonlep_bottom_part1} and \ref{Tab:nonlep_bottom_part2}.
	Comparisons with some recent works \cite{Geng:2018plk,Huber:2016xod} and the experimental results \cite{Olive:2016xmw} can
	be found in Tab.~\ref{Tab:comparison_nonlep_charm} and Tab.~\ref{Tab:comparison_nonlep_bottom}.
	
	\begin{table}
		\caption{Nonleptonic decays for charmed baryons. }
		\label{Tab:nonlep_charm} %
		\begin{tabular}{l|c|c|l|c|c}
			\hline 
			channels  & $\Gamma/{\rm GeV}$  & ${\cal B}$  & channels  & $\Gamma/{\rm GeV}$  & ${\cal B}$ \tabularnewline
			\hline 
			$\Lambda_{c}^{+}\to n\pi^{+}$  & $3.97\times10^{-15}$  & $1.21\times10^{-3}$  & $\Lambda_{c}^{+}\to n\rho^{+}$  & $1.33\times10^{-14}$  & $4.04\times10^{-3}$ \tabularnewline
			$\Lambda_{c}^{+}\to na_{1}^{+}$  & $9.88\times10^{-15}$  & $3.00\times10^{-3}$  & $\Lambda_{c}^{+}\to nK^{+}$  & $3.04\times10^{-16}$  & $9.23\times10^{-5}$ \tabularnewline
			$\Lambda_{c}^{+}\to nK^{*+}$  & $6.71\times10^{-16}$  & $2.04\times10^{-4}$  &  &  & \tabularnewline
			\hline 
			$\Lambda_{c}^{+}\to\Lambda\pi^{+}$  & $4.77\times10^{-14}$  & $1.45\times10^{-2}$  & $\Lambda_{c}^{+}\to\Lambda\rho^{+}$  & $1.39\times10^{-13}$  & $4.24\times10^{-2}$ \tabularnewline
			$\Lambda_{c}^{+}\to\Lambda K^{*+}$  & $6.47\times10^{-15}$  & $1.97\times10^{-3}$  & $\Lambda_{c}^{+}\to\Lambda K^{+}$  & $3.47\times10^{-15}$  & $1.05\times10^{-3}$ \tabularnewline
			\hline 
			$\Xi_{c}^{+}\to\Sigma^{0}\pi^{+}$  & $1.90\times10^{-15}$  & $1.28\times10^{-3}$  & $\Xi_{c}^{+}\to\Sigma^{0}\rho^{+}$  & $6.21\times10^{-15}$  & $4.17\times10^{-3}$ \tabularnewline
			$\Xi_{c}^{+}\to\Sigma^{0}a_{1}^{+}$  & $2.96\times10^{-15}$  & $1.99\times10^{-3}$  & $\Xi_{c}^{+}\to\Sigma^{0}K^{+}$  & $1.45\times10^{-16}$  & $9.74\times10^{-5}$ \tabularnewline
			$\Xi_{c}^{+}\to\Sigma^{0}K^{*+}$  & $3.08\times10^{-16}$  & $2.07\times10^{-4}$  &  &  & \tabularnewline
			\hline 
			$\Xi_{c}^{+}\to\Lambda\pi^{+}$  & $7.09\times10^{-16}$  & $4.76\times10^{-4}$  & $\Xi_{c}^{+}\to\Lambda\rho^{+}$  & $2.41\times10^{-15}$  & $1.62\times10^{-3}$ \tabularnewline
			$\Xi_{c}^{+}\to\Lambda a_{1}^{+}$  & $1.91\times10^{-15}$  & $1.29\times10^{-3}$  & $\Xi_{c}^{+}\to\Lambda K^{+}$  & $5.49\times10^{-17}$  & $3.69\times10^{-5}$ \tabularnewline
			$\Xi_{c}^{+}\to\Lambda K^{*+}$  & $1.23\times10^{-16}$  & $8.24\times10^{-5}$  &  &  & \tabularnewline
			\hline 
			$\Xi_{c}^{+}\to\Xi^{0}\pi^{+}$  & $7.30\times10^{-14}$  & $4.91\times10^{-2}$  & $\Xi_{c}^{+}\to\Xi^{0}\rho^{+}$  & $2.13\times10^{-13}$  & $1.43\times10^{-1}$ \tabularnewline
			$\Xi_{c}^{+}\to\Xi^{0}K^{*+}$  & $9.85\times10^{-15}$  & $6.62\times10^{-3}$  & $\Xi_{c}^{+}\to\Xi^{0}K^{+}$  & $5.33\times10^{-15}$  & $3.58\times10^{-3}$ \tabularnewline
			\hline 
			$\Xi_{c}^{0}\to\Sigma^{-}\pi^{+}$  & $3.80\times10^{-15}$  & $6.46\times10^{-4}$  & $\Xi_{c}^{0}\to\Sigma^{-}\rho^{+}$  & $1.24\times10^{-14}$  & $2.11\times10^{-3}$ \tabularnewline
			$\Xi_{c}^{0}\to\Sigma^{-}a_{1}^{+}$  & $5.85\times10^{-15}$  & $9.95\times10^{-4}$  & $\Xi_{c}^{0}\to\Sigma^{-}K^{+}$  & $2.90\times10^{-16}$  & $4.93\times10^{-5}$ \tabularnewline
			$\Xi_{c}^{0}\to\Sigma^{-}K^{*+}$  & $6.16\times10^{-16}$  & $1.05\times10^{-4}$  &  &  & \tabularnewline
			\hline 
			$\Xi_{c}^{0}\to\Xi^{-}\pi^{+}$  & $7.26\times10^{-14}$  & $1.24\times10^{-2}$  & $\Xi_{c}^{0}\to\Xi^{-}\rho^{+}$  & $2.11\times10^{-13}$  & $3.60\times10^{-2}$ \tabularnewline
			$\Xi_{c}^{0}\to\Xi^{-}K^{*+}$  & $9.73\times10^{-15}$  & $1.66\times10^{-3}$  & $\Xi_{c}^{0}\to\Xi^{-}K^{+}$  & $5.29\times10^{-15}$  & $9.01\times10^{-4}$ \tabularnewline
			\hline 
			$\Omega_{c}^{0}\to\Xi^{-}\pi^{+}$  & $1.68\times10^{-15}$  & $1.76\times10^{-4}$  & $\Omega_{c}^{0}\to\Xi^{-}\rho^{+}$  & $4.28\times10^{-15}$  & $4.49\times10^{-4}$ \tabularnewline
			$\Omega_{c}^{0}\to\Xi^{-}a_{1}^{+}$  & $2.51\times10^{-15}$  & $2.63\times10^{-4}$  & $\Omega_{c}^{0}\to\Xi^{-}K^{+}$  & $1.49\times10^{-16}$  & $1.57\times10^{-5}$ \tabularnewline
			$\Omega_{c}^{0}\to\Xi^{-}K^{*+}$  & $2.03\times10^{-16}$  & $2.13\times10^{-5}$  &  &  & \tabularnewline
			\hline 
		\end{tabular}
	\end{table}
	
	\begin{table}
		\caption{A comparison with some recent works for nonleptonic charmed decays. }
		\label{Tab:comparison_nonlep_charm} %
		\begin{tabular}{c|c|c|c}
			\hline 
			channel  & this work  & Ref.~\cite{Geng:2018plk}  & experiment~\cite{Olive:2016xmw}\tabularnewline
			\hline 
			$\Lambda_{c}^{+}\to\Lambda\pi^{+}$  & $1.45\times10^{-2}$  & - -  & $(1.30\pm0.07)\times10^{-2}$\tabularnewline
			$\Lambda_{c}^{+}\to\Lambda\rho^{+}$  & $4.24\times10^{-2}$  & - -  & $<6\times10^{-2}$\tabularnewline
			$\Lambda_{c}^{+}\to\Lambda K^{+}$  & $1.05\times10^{-3}$  & - -  & $(0.61\pm0.12)\times10^{-3}$\tabularnewline
			$\Xi_{c}^{+}\to\Xi^{0}\pi^{+}$  & $4.91\times10^{-2}$  & $(0.81\pm0.40)\times10^{-2}$  & - -\tabularnewline
			$\Xi_{c}^{0}\to\Xi^{-}\pi^{+}$  & $1.24\times10^{-2}$  & $(1.57\pm0.07)\times10^{-2}$  & - -\tabularnewline
			\hline 
		\end{tabular}
	\end{table}
	
	\begin{table}
		\caption{Nonleptonic decays for $\Lambda_{b}$ and $\Xi_{b}$. }
		\label{Tab:nonlep_bottom_part1} %
		\begin{tabular}{l|c|c|l|c|c}
			\hline 
			channels  & $\Gamma/{\rm GeV}$  & ${\cal B}$  & channels  & $\Gamma/{\rm GeV}$  & ${\cal B}$ \tabularnewline
			\hline 
			$\Lambda_{b}^{0}\to p\pi^{-}$  & $3.99\times10^{-18}$  & $8.90\times10^{-6}$  & $\Lambda_{b}^{0}\to p\rho^{-}$  & $1.17\times10^{-17}$  & $2.61\times10^{-5}$ \tabularnewline
			$\Lambda_{b}^{0}\to pa_{1}^{-}$  & $1.56\times10^{-17}$  & $3.48\times10^{-5}$  & $\Lambda_{b}^{0}\to pK^{-}$  & $3.22\times10^{-19}$  & $7.18\times10^{-7}$ \tabularnewline
			$\Lambda_{b}^{0}\to pK^{*-}$  & $6.02\times10^{-19}$  & $1.34\times10^{-6}$  & $\Lambda_{b}^{0}\to pD^{-}$  & $5.76\times10^{-19}$  & $1.28\times10^{-6}$ \tabularnewline
			$\Lambda_{b}^{0}\to pD^{*-}$  & $8.95\times10^{-19}$  & $1.99\times10^{-6}$  & $\Lambda_{b}^{0}\to pD_{s}^{-}$  & $1.54\times10^{-17}$  & $3.44\times10^{-5}$ \tabularnewline
			$\Lambda_{b}^{0}\to pD_{s}^{*-}$  & $2.19\times10^{-17}$  & $4.88\times10^{-5}$  &  &  & \tabularnewline
			\hline 
			$\Lambda_{b}^{0}\to\Lambda_{c}^{+}\pi^{-}$  & $3.83\times10^{-15}$  & $8.53\times10^{-3}$  & $\Lambda_{b}^{0}\to\Lambda_{c}^{+}\rho^{-}$  & $1.09\times10^{-14}$  & $2.44\times10^{-2}$ \tabularnewline
			$\Lambda_{b}^{0}\to\Lambda_{c}^{+}a_{1}^{-}$  & $1.40\times10^{-14}$  & $3.12\times10^{-2}$  & $\Lambda_{b}^{0}\to\Lambda_{c}^{+}K^{-}$  & $3.04\times10^{-16}$  & $6.78\times10^{-4}$ \tabularnewline
			$\Lambda_{b}^{0}\to\Lambda_{c}^{+}K^{*-}$  & $5.59\times10^{-16}$  & $1.24\times10^{-3}$  & $\Lambda_{b}^{0}\to\Lambda_{c}^{+}D^{-}$  & $4.26\times10^{-16}$  & $9.49\times10^{-4}$ \tabularnewline
			$\Lambda_{b}^{0}\to\Lambda_{c}^{+}D^{*-}$  & $6.90\times10^{-16}$  & $1.54\times10^{-3}$  & $\Lambda_{b}^{0}\to\Lambda_{c}^{+}D_{s}^{-}$  & $1.10\times10^{-14}$  & $2.46\times10^{-2}$ \tabularnewline
			$\Lambda_{b}^{0}\to\Lambda_{c}^{+}D_{s}^{*-}$  & $1.64\times10^{-14}$  & $3.65\times10^{-2}$  &  &  & \tabularnewline
			\hline 
			$\Xi_{b}^{0}\to\Sigma^{+}\pi^{-}$  & $3.55\times10^{-18}$  & $7.91\times10^{-6}$  & $\Xi_{b}^{0}\to\Sigma^{+}\rho^{-}$  & $1.05\times10^{-17}$  & $2.33\times10^{-5}$ \tabularnewline
			$\Xi_{b}^{0}\to\Sigma^{+}a_{1}^{-}$  & $1.41\times10^{-17}$  & $3.13\times10^{-5}$  & $\Xi_{b}^{0}\to\Sigma^{+}K^{-}$  & $2.87\times10^{-19}$  & $6.40\times10^{-7}$ \tabularnewline
			$\Xi_{b}^{0}\to\Sigma^{+}K^{*-}$  & $5.39\times10^{-19}$  & $1.20\times10^{-6}$  & $\Xi_{b}^{0}\to\Sigma^{+}D^{-}$  & $5.30\times10^{-19}$  & $1.18\times10^{-6}$ \tabularnewline
			$\Xi_{b}^{0}\to\Sigma^{+}D^{*-}$  & $8.23\times10^{-19}$  & $1.83\times10^{-6}$  & $\Xi_{b}^{0}\to\Sigma^{+}D_{s}^{-}$  & $1.42\times10^{-17}$  & $3.17\times10^{-5}$ \tabularnewline
			$\Xi_{b}^{0}\to\Sigma^{+}D_{s}^{*-}$  & $2.02\times10^{-17}$  & $4.50\times10^{-5}$  &  &  & \tabularnewline
			\hline 
			$\Xi_{b}^{0}\to\Xi_{c}^{+}\pi^{-}$  & $3.76\times10^{-15}$  & $8.37\times10^{-3}$  & $\Xi_{b}^{0}\to\Xi_{c}^{+}\rho^{-}$  & $1.08\times10^{-14}$  & $2.40\times10^{-2}$ \tabularnewline
			$\Xi_{b}^{0}\to\Xi_{c}^{+}a_{1}^{-}$  & $1.38\times10^{-14}$  & $3.08\times10^{-2}$  & $\Xi_{b}^{0}\to\Xi_{c}^{+}K^{-}$  & $3.00\times10^{-16}$  & $6.67\times10^{-4}$ \tabularnewline
			$\Xi_{b}^{0}\to\Xi_{c}^{+}K^{*-}$  & $5.51\times10^{-16}$  & $1.23\times10^{-3}$  & $\Xi_{b}^{0}\to\Xi_{c}^{+}D^{-}$  & $4.26\times10^{-16}$  & $9.49\times10^{-4}$ \tabularnewline
			$\Xi_{b}^{0}\to\Xi_{c}^{+}D^{*-}$  & $6.90\times10^{-16}$  & $1.54\times10^{-3}$  & $\Xi_{b}^{0}\to\Xi_{c}^{+}D_{s}^{-}$  & $1.11\times10^{-14}$  & $2.46\times10^{-2}$ \tabularnewline
			$\Xi_{b}^{0}\to\Xi_{c}^{+}D_{s}^{*-}$  & $1.64\times10^{-14}$  & $3.65\times10^{-2}$  &  &  & \tabularnewline
			\hline 
			$\Xi_{b}^{-}\to\Sigma^{0}\pi^{-}$  & $1.78\times10^{-18}$  & $4.22\times10^{-6}$  & $\Xi_{b}^{-}\to\Sigma^{0}\rho^{-}$  & $5.23\times10^{-18}$  & $1.24\times10^{-5}$ \tabularnewline
			$\Xi_{b}^{-}\to\Sigma^{0}a_{1}^{-}$  & $7.03\times10^{-18}$  & $1.67\times10^{-5}$  & $\Xi_{b}^{-}\to\Sigma^{0}K^{-}$  & $1.44\times10^{-19}$  & $3.41\times10^{-7}$ \tabularnewline
			$\Xi_{b}^{-}\to\Sigma^{0}K^{*-}$  & $2.70\times10^{-19}$  & $6.40\times10^{-7}$  & $\Xi_{b}^{-}\to\Sigma^{0}D^{-}$  & $2.65\times10^{-19}$  & $6.29\times10^{-7}$ \tabularnewline
			$\Xi_{b}^{-}\to\Sigma^{0}D^{*-}$  & $4.12\times10^{-19}$  & $9.76\times10^{-7}$  & $\Xi_{b}^{-}\to\Sigma^{0}D_{s}^{-}$  & $7.13\times10^{-18}$  & $1.69\times10^{-5}$ \tabularnewline
			$\Xi_{b}^{-}\to\Sigma^{0}D_{s}^{*-}$  & $1.01\times10^{-17}$  & $2.40\times10^{-5}$  &  &  & \tabularnewline
			\hline 
			$\Xi_{b}^{-}\to\Lambda\pi^{-}$  & $6.03\times10^{-19}$  & $1.43\times10^{-6}$  & $\Xi_{b}^{-}\to\Lambda\rho^{-}$  & $1.77\times10^{-18}$  & $4.21\times10^{-6}$ \tabularnewline
			$\Xi_{b}^{-}\to\Lambda a_{1}^{-}$  & $2.39\times10^{-18}$  & $5.66\times10^{-6}$  & $\Xi_{b}^{-}\to\Lambda K^{-}$  & $4.88\times10^{-20}$  & $1.16\times10^{-7}$ \tabularnewline
			$\Xi_{b}^{-}\to\Lambda K^{*-}$  & $9.15\times10^{-20}$  & $2.17\times10^{-7}$  & $\Xi_{b}^{-}\to\Lambda D^{-}$  & $9.02\times10^{-20}$  & $2.14\times10^{-7}$ \tabularnewline
			$\Xi_{b}^{-}\to\Lambda D^{*-}$  & $1.40\times10^{-19}$  & $3.32\times10^{-7}$  & $\Xi_{b}^{-}\to\Lambda D_{s}^{-}$  & $2.43\times10^{-18}$  & $5.75\times10^{-6}$ \tabularnewline
			$\Xi_{b}^{-}\to\Lambda D_{s}^{*-}$  & $3.44\times10^{-18}$  & $8.16\times10^{-6}$  &  &  & \tabularnewline
			\hline 
			$\Xi_{b}^{-}\to\Xi_{c}^{0}\pi^{-}$  & $3.76\times10^{-15}$  & $8.93\times10^{-3}$  & $\Xi_{b}^{-}\to\Xi_{c}^{0}\rho^{-}$  & $1.08\times10^{-14}$  & $2.56\times10^{-2}$ \tabularnewline
			$\Xi_{b}^{-}\to\Xi_{c}^{0}a_{1}^{-}$  & $1.38\times10^{-14}$  & $3.28\times10^{-2}$  & $\Xi_{b}^{-}\to\Xi_{c}^{0}K^{-}$  & $3.00\times10^{-16}$  & $7.11\times10^{-4}$ \tabularnewline
			$\Xi_{b}^{-}\to\Xi_{c}^{0}K^{*-}$  & $5.51\times10^{-16}$  & $1.31\times10^{-3}$  & $\Xi_{b}^{-}\to\Xi_{c}^{0}D^{-}$  & $4.27\times10^{-16}$  & $1.01\times10^{-3}$ \tabularnewline
			$\Xi_{b}^{-}\to\Xi_{c}^{0}D^{*-}$  & $6.91\times10^{-16}$  & $1.64\times10^{-3}$  & $\Xi_{b}^{-}\to\Xi_{c}^{0}D_{s}^{-}$  & $1.11\times10^{-14}$  & $2.62\times10^{-2}$ \tabularnewline
			$\Xi_{b}^{-}\to\Xi_{c}^{0}D_{s}^{*-}$  & $1.64\times10^{-14}$  & $3.90\times10^{-2}$  &  &  & \tabularnewline
			\hline 
		\end{tabular}
	\end{table}
	
	\begin{table}
		\caption{Nonleptonic decays for $\Omega_{b}$. }
		\label{Tab:nonlep_bottom_part2} %
		\begin{tabular}{l|c|c|l|c|c}
			\hline 
			channels  & $\Gamma/{\rm GeV}$  & ${\cal B}$  & channels  & $\Gamma/{\rm GeV}$  & ${\cal B}$ \tabularnewline
			\hline 
			$\Omega_{b}^{-}\to\Xi^{0}\pi^{-}$  & $6.05\times10^{-19}$  & $1.44\times10^{-6}$  & $\Omega_{b}^{-}\to\Xi^{0}\rho^{-}$  & $1.73\times10^{-18}$  & $4.13\times10^{-6}$ \tabularnewline
			$\Omega_{b}^{-}\to\Xi^{0}a_{1}^{-}$  & $2.23\times10^{-18}$  & $5.33\times10^{-6}$  & $\Omega_{b}^{-}\to\Xi^{0}K^{-}$  & $5.00\times10^{-20}$  & $1.19\times10^{-7}$ \tabularnewline
			$\Omega_{b}^{-}\to\Xi^{0}K^{*-}$  & $8.86\times10^{-20}$  & $2.11\times10^{-7}$  & $\Omega_{b}^{-}\to\Xi^{0}D^{-}$  & $1.21\times10^{-19}$  & $2.88\times10^{-7}$ \tabularnewline
			$\Omega_{b}^{-}\to\Xi^{0}D^{*-}$  & $1.13\times10^{-19}$  & $2.69\times10^{-7}$  & $\Omega_{b}^{-}\to\Xi^{0}D_{s}^{-}$  & $3.32\times10^{-18}$  & $7.91\times10^{-6}$ \tabularnewline
			$\Omega_{b}^{-}\to\Xi^{0}D_{s}^{*-}$  & $2.68\times10^{-18}$  & $6.40\times10^{-6}$  &  &  & \tabularnewline
			\hline 
			$\Omega_{b}^{-}\to\Omega_{c}^{0}\pi^{-}$  & $1.68\times10^{-15}$  & $4.00\times10^{-3}$  & $\Omega_{b}^{-}\to\Omega_{c}^{0}\rho^{-}$  & $4.54\times10^{-15}$  & $1.08\times10^{-2}$ \tabularnewline
			$\Omega_{b}^{-}\to\Omega_{c}^{0}a_{1}^{-}$  & $5.37\times10^{-15}$  & $1.28\times10^{-2}$  & $\Omega_{b}^{-}\to\Omega_{c}^{0}K^{-}$  & $1.36\times10^{-16}$  & $3.26\times10^{-4}$ \tabularnewline
			$\Omega_{b}^{-}\to\Omega_{c}^{0}K^{*-}$  & $2.28\times10^{-16}$  & $5.44\times10^{-4}$  & $\Omega_{b}^{-}\to\Omega_{c}^{0}D^{-}$  & $2.66\times10^{-16}$  & $6.36\times10^{-4}$ \tabularnewline
			$\Omega_{b}^{-}\to\Omega_{c}^{0}D^{*-}$  & $2.14\times10^{-16}$  & $5.11\times10^{-4}$  & $\Omega_{b}^{-}\to\Omega_{c}^{0}D_{s}^{-}$  & $7.17\times10^{-15}$  & $1.71\times10^{-2}$ \tabularnewline
			$\Omega_{b}^{-}\to\Omega_{c}^{0}D_{s}^{*-}$  & $4.90\times10^{-15}$  & $1.17\times10^{-2}$  &  &  & \tabularnewline
			\hline 
		\end{tabular}
	\end{table}
	
	\begin{table}
		\caption{A comparison with some recent works for nonleptonic bottomed decays. }
		\label{Tab:comparison_nonlep_bottom} %
		\begin{tabular}{c|c|c|c}
			\hline 
			channel  & this work  & Ref.~\cite{Huber:2016xod} & experiment~\cite{Olive:2016xmw}\tabularnewline
			\hline 
			$\Lambda_{b}^{0}\to\Lambda_{c}^{+}\pi^{-}$  & $8.53\times10^{-3}$  & $(2.85\pm0.54)\times10^{-3}$  & $(4.9\pm0.4)\times10^{-3}$\tabularnewline
			$\Lambda_{b}^{0}\to\Lambda_{c}^{+}\rho^{-}$  & $2.44\times10^{-2}$  & $(0.817\pm0.147)\times10^{-2}$  & - -\tabularnewline
			$\Lambda_{b}^{0}\to\Lambda_{c}^{+}a_{1}^{-}$  & $3.12\times10^{-2}$  & $(1.047\pm0.178)\times10^{-2}$  & - -\tabularnewline
			$\Lambda_{b}^{0}\to\Lambda_{c}^{+}K^{-}$  & $6.78\times10^{-4}$ & $(2.21\pm0.40)\times10^{-4}$  & $(3.59\pm0.30)\times10^{-4}$ \tabularnewline
			$\Lambda_{b}^{0}\to\Lambda_{c}^{+}K^{*-}$  & $1.24\times10^{-3}$  & $(0.422\pm0.075)\times10^{-3}$  & - -\tabularnewline
			$\Lambda_{b}^{0}\to\Lambda_{c}^{+}D^{-}$  & $9.49\times10^{-4}$  & - - & $(4.6\pm0.6)\times10^{-4}$ \tabularnewline
			$\Lambda_{b}^{0}\to\Lambda_{c}^{+}D_{s}^{-}$  & $2.46\times10^{-2}$  & - - & $(1.10\pm0.10)\times10^{-2}$ \tabularnewline
			$\Lambda_{b}^{0}\to p\pi^{-}$  & $8.90\times10^{-6}$  & - - & $(4.2\pm0.8)\times10^{-6}$\tabularnewline
			$\Lambda_{b}^{0}\to pK^{-}$  & $7.18\times10^{-7}$  & - - & $(51\pm10)\times10^{-7}$\tabularnewline
			$\Lambda_{b}^{0}\to pD_{s}^{-}$  & $3.44\times10^{-5}$  & - - & $<48\times10^{-5}$\tabularnewline
			\hline 
		\end{tabular}
	\end{table}

	\subsection{SU(3) analysis for semi-leptonic decays}
	
	From the overlapping factors above, we would expect the following
	relations 
\begin{eqnarray}
&  & \frac{2\Gamma(\Lambda_{c}^{+}\to ne^{+}\nu_{e})}{|V_{cd}|^{2}}=\frac{3\Gamma(\Lambda_{c}^{+}\to\Lambda e^{+}\nu_{e})}{|V_{cs}|^{2}}\nonumber \\
& = & \frac{4\Gamma(\Xi_{c}^{+}\to\Sigma^{0}e^{+}\nu_{e})}{|V_{cd}|^{2}}=\frac{12\Gamma(\Xi_{c}^{+}\to\Lambda e^{+}\nu_{e})}{|V_{cd}|^{2}}=\frac{2\Gamma(\Xi_{c}^{+}\to\Xi^{0}e^{+}\nu_{e})}{|V_{cs}|^{2}}\nonumber \\
& = & \frac{2\Gamma(\Xi_{c}^{0}\to\Sigma^{-}e^{+}\nu_{e})}{|V_{cd}|^{2}}=\frac{2\Gamma(\Xi_{c}^{0}\to\Xi^{-}e^{+}\nu_{e})}{|V_{cs}|^{2}}
\end{eqnarray}
	for c-baryon sector and 
\begin{eqnarray}
&  & \Gamma(\Lambda_{b}^{0}\to pe^{-}\bar{\nu}_{e})=\Gamma(\Xi_{b}^{0}\to\Sigma^{+}e^{-}\bar{\nu}_{e})=2\Gamma(\Xi_{b}^{-}\to\Sigma^{0}e^{-}\bar{\nu}_{e})=6\Gamma(\Xi_{b}^{-}\to\Lambda e^{-}\bar{\nu}_{e}),\nonumber \\
&  & \Gamma(\Lambda_{b}^{0}\to\Lambda_{c}^{+}e^{-}\bar{\nu}_{e})=\Gamma(\Xi_{b}^{0}\to\Xi_{c}^{+}e^{-}\bar{\nu}_{e})=\Gamma(\Xi_{b}^{-}\to\Xi_{c}^{0}e^{-}\bar{\nu}_{e})
\end{eqnarray}
	for b-baryon sector, if the flavor SU(3) symmetry is respected. These
	relations for the charmed baryons are consistent with those in Refs.~\cite{Lu:2016ogy,Geng:2017mxn}, while the ones for the bottomed baryons, as far as we know, are
	first derived by this work.
	
	In the following, we will investigate the SU(3) symmetry breaking
	effects. The corresponding results are collected in Tabs.~\ref{Tab:SU3_breaking_charm}
	and \ref{Tab:SU3_breaking_bottom}. Take $\Lambda_{c}^{+}\to ne^{+}\nu_{e}$
	and $\Lambda_{c}^{+}\to\Lambda e^{+}\nu_{e}$ as examples. we can
	see that: 
	\begin{itemize}
		\item If we have considered the differences of CKM and the overlapping factors
		between these two channels but take all the other parameters as the
		same, we get the precise SU(3) symmetry prediction $\Gamma(\Lambda_{c}^{+}\to ne^{+}\nu_{e})/(\frac{1}{2}|V_{cd}|^{2})=\Gamma(\Lambda_{c}^{+}\to\Lambda e^{+}\nu_{e})/(\frac{1}{3}|V_{cs}|^{2})$.
		This prediction is also obtained in Refs.~\cite{Geng:2017mxn,Lu:2016ogy}. 
		\item If we consider only the difference of daughter baryon mass but take
		all the other parameters as the same, we get a ratio 0.538. It
		means that SU(3) symmetry is broken by about 50\% between these two
		modes. The more accurate number is 35\% (see Tab.~\ref{Tab:SU3_breaking_charm}),
		when   all the other relevant impacts are taken into account. 
	\end{itemize}
	We can see from Tabs.~\ref{Tab:SU3_breaking_charm} and \ref{Tab:SU3_breaking_bottom}: 
	\begin{itemize}
		\item The SU(3) symmetry breaking is sizable for c-baryon decays while it
		is small for the b-baryon decays. This can be understood due to a
		much smaller phase space in c-baryon decays, and thus the decay width
		significantly depends on  the mass differences of the baryons in the initial and
		final states. 
		\item SU(3) symmetry is broken more severely in the $c\to s$ processes
		than in the $c\to d$ processes because of the larger mass of $s$
		quark than $u$ and $d$ quarks. The typical value of SU(3) symmetry
		breaking for $c\to s$ processes is 35\% while that for $c\to d$
		processes is 15\%. 
	\end{itemize}
	\begin{table}
		\caption{Quantitative predictions of SU(3) breaking for semi-leptonic charmed
			decays. }
		\label{Tab:SU3_breaking_charm} %
		\begin{tabular}{l|c|c|c}
			\hline 
			channels  & $\Gamma/{\rm GeV}$ (LFQM)  & $\Gamma/{\rm GeV}$ (SU(3))  & $|{\rm LFQM}-{\rm SU(3)}|/{\rm SU(3)}$\tabularnewline
			\hline 
			$\Lambda_{c}^{+}\to ne^{+}\nu_{e}$  & $6.62\times10^{-15}$  & $6.62\times10^{-15}$  & - -\tabularnewline
			$\Lambda_{c}^{+}\to\Lambda e^{+}\nu_{e}$  & $5.36\times10^{-14}$  & $8.27\times10^{-14}$  & $35\%$\tabularnewline
			$\Xi_{c}^{+}\to\Sigma^{0}e^{+}\nu_{e}$  & $2.79\times10^{-15}$  & $3.31\times10^{-15}$  & $16\%$\tabularnewline
			$\Xi_{c}^{+}\to\Lambda e^{+}\nu_{e}$  & $1.22\times10^{-15}$  & $1.10\times10^{-15}$  & $11\%$\tabularnewline
			$\Xi_{c}^{+}\to\Xi^{0}e^{+}\nu_{e}$  & $8.03\times10^{-14}$  & $1.24\times10^{-13}$  & $35\%$\tabularnewline
			$\Xi_{c}^{0}\to\Sigma^{-}e^{+}\nu_{e}$  & $5.57\times10^{-15}$  & $6.62\times10^{-15}$  & $16\%$\tabularnewline
			$\Xi_{c}^{0}\to\Xi^{-}e^{+}\nu_{e}$  & $7.91\times10^{-14}$  & $1.24\times10^{-13}$  & $36\%$\tabularnewline
			\hline 
		\end{tabular}
	\end{table}
	
	\begin{table}
		\caption{Same as Tab.~\ref{Tab:SU3_breaking_charm} but for bottomed case. }
		\label{Tab:SU3_breaking_bottom} %
		\begin{tabular}{c|c|c|c}
			\hline 
			channels  & $\Gamma/{\rm GeV}$ (LFQM)  & $\Gamma/{\rm GeV}$ (SU(3))  & $|{\rm LFQM}-{\rm SU(3)}|/{\rm SU(3)}$\tabularnewline
			\hline 
			$\Lambda_{b}^{0}\to pe^{-}\bar{\nu}_{e}$  & $1.41\times10^{-16}$  & $1.41\times10^{-16}$  & - -\tabularnewline
			$\Xi_{b}^{0}\to\Sigma^{+}e^{-}\bar{\nu}_{e}$  & $1.27\times10^{-16}$  & $1.41\times10^{-16}$  & $9.9\%$\tabularnewline
			$\Xi_{b}^{-}\to\Sigma^{0}e^{-}\bar{\nu}_{e}$  & $6.37\times10^{-17}$  & $7.05\times10^{-17}$  & $9.6\%$\tabularnewline
			$\Xi_{b}^{-}\to\Lambda e^{-}\bar{\nu}_{e}$  & $2.29\times10^{-17}$  & $2.35\times10^{-17}$  & $2.6\%$\tabularnewline
			\hline 
			$\Lambda_{b}^{0}\to\Lambda_{c}^{+}e^{-}\bar{\nu}_{e}$  & $3.96\times10^{-14}$  & $3.96\times10^{-14}$  & - -\tabularnewline
			$\Xi_{b}^{0}\to\Xi_{c}^{+}e^{-}\bar{\nu}_{e}$  & $3.97\times10^{-14}$  & $3.96\times10^{-14}$  & $0.25\%$\tabularnewline
			$\Xi_{b}^{-}\to\Xi_{c}^{0}e^{-}\bar{\nu}_{e}$  & $3.97\times10^{-14}$  & $3.96\times10^{-14}$  & $0.25\%$\tabularnewline
			\hline 
		\end{tabular}
	\end{table}

\subsection{Uncertainties}

In this subsection, we will look more carefully at the dependence of our results on the model parameters. Taking $\Lambda_{c}^{+}\to \Lambda$ transition as an example. Varying the model parameters $m_{di}=m_{[ud]}$, $\beta_{i}=\beta_{c[ud]}$ and $\beta_{f}=\beta_{s[ud]}$ by 10\% respectively, we arrive at the following error estimates for form factors
\begin{eqnarray}
\label{eq:err_ff}
	f_{1}^{\Lambda_{c}^{+}\to\Lambda}(0) & = & 0.810\pm0.007\pm0.008\pm0.033,\nonumber \\
	f_{2}^{\Lambda_{c}^{+}\to\Lambda}(0) & = & -0.384\pm0.008\pm0.008\pm0.015,\nonumber \\
	g_{1}^{\Lambda_{c}^{+}\to\Lambda}(0) & = & 0.705\pm0.006\pm0.015\pm0.022,\nonumber \\
	g_{2}^{\Lambda_{c}^{+}\to\Lambda}(0) & = & -0.060\pm0.003\pm0.041\pm0.041
\end{eqnarray}
and for decay widths
\begin{eqnarray}
\label{eq:err_decay_width}
	\Gamma(\Lambda_{c}^{+}\to\Lambda e^{+}\nu_{e}) & = & (5.36\pm0.03\pm0.10\pm0.07)\times10^{-14},\nonumber \\
	\Gamma(\Lambda_{c}^{+}\to\Lambda\pi^{+}) & = & (4.77\pm0.08\pm0.14\pm0.34)\times10^{-14},\nonumber \\
	\Gamma(\Lambda_{c}^{+}\to\Lambda\rho^{+}) & = & (1.39\pm0.01\pm0.02\pm0.02)\times10^{-13},\nonumber \\
	\Gamma(\Lambda_{c}^{+}\to\Lambda K^{+}) & = & (3.47\pm0.04\pm0.11\pm0.22)\times10^{-15},\nonumber \\
	\Gamma(\Lambda_{c}^{+}\to\Lambda K^{*+}) & = & (6.47\pm0.02\pm0.10\pm0.19)\times10^{-15}.
\end{eqnarray}
Some comments are given in order.
\begin{itemize}
\item All the form factors are not very sensitive to the diquark mass $m_{di}$.
\item $g_{2}$ is one order of magnitude smaller than the other form factors, and it is sensitive to $\beta_{i}$ and $\beta_{f}$, while $f_{1}$, $f_{2}$ and $g_{1}$ are still not very sensitive to these shape parameters.
\item It can be seen from Eq. (\ref{eq:err_decay_width}) that, these decay widths are not sensitive to the model parameters. There exists at most about 10\% deviation in these decay widths.
\end{itemize}

	\section{Conclusion}
	
	In this work, we have calculated the transition form factors of the
	singly heavy baryons using the light-front approach under the diquark
	picture. These form factors are then used to predict semi-leptonic
	and nonleptonic decays of singly heavy baryons. Most of our results
	are comparable to the available experimental data and other theoretical
	results. We have also derived the overlapping factors that can be
	used to reproduce the SU(3) predictions on semi-leptonic decays. Using
	the calculated form factors, we pointed out that the SU(3) symmetry
	breaking is sizable in the charmed baryon decays while in the bottomed
	case, the SU(3) symmetry breaking is small. Most of the results in this work can be examined at experimental facilities at BEPCII, LHC or BELLEII. 
	
	\section*{Acknowledgements}
	
	The author is grateful to Prof.~Wei Wang for   valuable  discussions and constant encouragements. 
	This work is supported in part by National Natural Science Foundation
	of China under Grant Nos.~11575110, 11655002, 11735010, Natural Science
	Foundation of Shanghai under Grant No.~15DZ2272100.
	

	\appendix
	
	\section{Wave fuctions in initial and final states}
	
	\label{app:wave_functions}
	
	\subsection{Wave functions in the standard flavor-spin basis}
	
	The wave functions in the flavor space can also be found in Ref.~\cite{Chau:1995gk}.
	The wave functions of the singly heavy baryons in the initial states
	in the standard flavor-spin basis are given as follows.
	
	For ${\cal B}_{cqq}^{\boldsymbol{6}}$ ($\Sigma_{c}^{++,0}$ and $\Omega_{c}^{0}$), we have 
	\begin{equation}
	|{\cal B}_{cqq}^{\boldsymbol{6}},\uparrow\rangle=(qqc)\left(\frac{1}{\sqrt{6}}(\uparrow\downarrow\uparrow+\downarrow\uparrow\uparrow-2\uparrow\uparrow\downarrow)\right),
	\end{equation}
	where $q=u,d,s$ for $\Sigma_{c}^{++,0}$ and $\Omega_{c}^{0}$, respectively.
	
	For ${\cal B}_{cq_{1}q_{2}}^{\boldsymbol{6}}$ ($\Sigma_{c}^{+}$
	and $\Xi_{c}^{\prime+,\prime0}$), we have
	\begin{equation}
	|{\cal B}_{cq_{1}q_{2}}^{\boldsymbol{6}},\uparrow\rangle=\left(\frac{1}{\sqrt{2}}(q_{1}q_{2}+q_{2}q_{1})c\right)\left(\frac{1}{\sqrt{6}}(\uparrow\downarrow\uparrow+\downarrow\uparrow\uparrow-2\uparrow\uparrow\downarrow)\right),
	\end{equation}
	where $(q_{1},q_{2})=(u,d),(u,s),(d,s)$ for $\Sigma_{c}^{+}$ and
	$\Xi_{c}^{\prime+,\prime0}$, respectively.
	
	For ${\cal B}_{cq_{1}q_{2}}^{\bar{\boldsymbol{3}}}$ ($\Lambda_{c}^{+}$
	and $\Xi_{c}^{+,0}$),  we have 
	\begin{equation}
	|{\cal B}_{cq_{1}q_{2}}^{\bar{\boldsymbol{3}}},\uparrow\rangle=\left(\frac{1}{\sqrt{2}}(q_{1}q_{2}-q_{2}q_{1})c\right)\left(\frac{1}{\sqrt{2}}(\uparrow\downarrow\uparrow-\downarrow\uparrow\uparrow)\right),
	\end{equation}
	where $(q_{1},q_{2})=(u,d),(u,s),(d,s)$ for $\Lambda_{c}^{+}$ and
	$\Xi_{c}^{+,0}$, respectively.
	
	The wave functions of the baryon octet in the final states in the
	standard flavor-spin basis are given as follows.
	
	For ${\cal B}_{q_{1}q_{1}q_{2}}$ ($p$, $n$, $\Sigma^{+,-}$, $\Xi^{0,-}$), we have 
	\begin{eqnarray}
	|{\cal B}_{q_{1}q_{1}q_{2}},\uparrow\rangle & = & \frac{1}{\sqrt{2}}\Bigg\{\left(\frac{1}{\sqrt{6}}(q_{1}q_{2}q_{1}+q_{2}q_{1}q_{1}-2q_{1}q_{1}q_{2})\right)\left(\frac{1}{\sqrt{6}}(\uparrow\downarrow\uparrow+\downarrow\uparrow\uparrow-2\uparrow\uparrow\downarrow)\right)\nonumber \\
	&  & +\left(\frac{1}{\sqrt{2}}(q_{1}q_{2}q_{1}-q_{2}q_{1}q_{1})\right)\left(\frac{1}{\sqrt{2}}(\uparrow\downarrow\uparrow-\downarrow\uparrow\uparrow)\right)\Bigg\},
	\end{eqnarray}
	where $(q_{1},q_{2})=(u,d),(d,u),(u,s),(d,s),(s,u),(s,d)$ for $p$,
	$n$, $\Sigma^{+,-}$, $\Xi^{0,-}$, respectively.
	
	For $\Sigma^{0}$ and $\Lambda$,  we have
	\begin{eqnarray}
	|\Sigma^{0},\uparrow\rangle & = & \frac{1}{\sqrt{2}}\Bigg\{\left(\frac{1}{\sqrt{12}}(sdu+dsu+sud+usd-2dus-2uds)\right)\left(\frac{1}{\sqrt{6}}(\uparrow\downarrow\uparrow+\downarrow\uparrow\uparrow-2\uparrow\uparrow\downarrow)\right)\nonumber \\
	&  & +\left(\frac{1}{2}(-sdu+dsu-sud+usd)\right)\left(\frac{1}{\sqrt{2}}(\uparrow\downarrow\uparrow-\downarrow\uparrow\uparrow)\right)\Bigg\},\\
	|\Lambda,\uparrow\rangle & = & \frac{1}{\sqrt{2}}\Bigg\{\left(\frac{1}{2}(sdu+dsu-sud-usd)\right)\left(\frac{1}{\sqrt{6}}(\uparrow\downarrow\uparrow+\downarrow\uparrow\uparrow-2\uparrow\uparrow\downarrow)\right)\nonumber \\
	&  & +\left(\frac{1}{\sqrt{12}}(sdu-dsu-sud+usd-2dus+2uds)\right)\left(\frac{1}{\sqrt{2}}(\uparrow\downarrow\uparrow-\downarrow\uparrow\uparrow)\right)\Bigg\}.
	\end{eqnarray}

	\subsection{Wave functions in the diquark basis}
	
	From the coupling of two angular momenta $j_{1}=1$ and $j_{2}=\frac{1}{2}$,
	we know that 
	\[
	|J=\frac{1}{2},M=\frac{1}{2}\rangle=\sqrt{\frac{2}{3}}|m_{1}=1,m_{2}=-\frac{1}{2}\rangle-\sqrt{\frac{1}{3}}|m_{1}=0,m_{2}=\frac{1}{2}\rangle.
	\]
	So, it is natural to define the baryon state with an axial-vector
	diquark as follows. 
	\begin{equation}
	|q_{1}(q_{2}q_{3})_{A},\uparrow\rangle\equiv\sqrt{\frac{2}{3}}q_{1}\downarrow(q_{2}q_{3})_{11}-\sqrt{\frac{1}{3}}q_{1}\uparrow(q_{2}q_{3})_{10},
	\end{equation}
	where $(q_{2}q_{3})_{11}=(q_{2}q_{3})(\uparrow\uparrow)$ and $(q_{2}q_{3})_{10}=(q_{2}q_{3})\left(\frac{1}{\sqrt{2}}(\uparrow\downarrow+\downarrow\uparrow)\right)$.
	Meanwhile the baryon state with a scalar diquark can be defined as
	\begin{equation}
	|q_{1}(q_{2}q_{3})_{S},\uparrow\rangle\equiv q_{1}\uparrow(q_{2}q_{3})_{S},
	\end{equation}
	where $(q_{2}q_{3})_{S}=(q_{2}q_{3})_{00}=(q_{2}q_{3})\left(\frac{1}{\sqrt{2}}(\uparrow\downarrow-\downarrow\uparrow)\right)$.
	
	One can prove the following equations
	\begin{eqnarray}
	q_{1}q_{2}q_{3}\left(\frac{1}{\sqrt{2}}(\uparrow\downarrow\uparrow-\downarrow\uparrow\uparrow)\right) & = & -\frac{1}{2}|q_{1}(q_{2}q_{3})_{S},\uparrow\rangle-\frac{\sqrt{3}}{2}|q_{1}(q_{2}q_{3})_{A},\uparrow\rangle,\\
	q_{1}q_{2}q_{3}\left(\frac{1}{\sqrt{6}}(\uparrow\downarrow\uparrow+\downarrow\uparrow\uparrow-2\uparrow\uparrow\downarrow)\right) & = & -\frac{\sqrt{3}}{2}|q_{1}(q_{2}q_{3})_{S},\uparrow\rangle+\frac{1}{2}|q_{1}(q_{2}q_{3})_{A},\uparrow\rangle.
	\end{eqnarray}
	
	Equipped with above expressions, the baryon wave functions in the
	initial and final states in the diquark basis can be derived as follows.
	
	For ${\cal B}_{cqq}^{\boldsymbol{6}}$ ($\Sigma_{c}^{++,0}$ and $\Omega_{c}^{0}$), we have
	\begin{equation}
	{\cal B}_{cqq}^{\boldsymbol{6}}=-c(qq)_{A},
	\end{equation}
	where $q=u,d,s$ for $\Sigma_{c}^{++,0}$ and $\Omega_{c}^{0}$, respectively.
	
	For ${\cal B}_{cq_{1}q_{2}}^{\boldsymbol{6}}$ ($\Sigma_{c}^{+}$
	and $\Xi_{c}^{\prime+,\prime0}$), we have
	\begin{equation}
	{\cal B}_{cq_{1}q_{2}}^{\boldsymbol{6}}=\frac{1}{\sqrt{2}}(-c(q_{1}q_{2})_{A}-c(q_{2}q_{1})_{A}),
	\end{equation}
	where $(q_{1},q_{2})=(u,d),(u,s),(d,s)$ for $\Sigma_{c}^{+}$ and
	$\Xi_{c}^{\prime+,\prime0}$, respectively.
	
	For ${\cal B}_{cq_{1}q_{2}}^{\bar{\boldsymbol{3}}}$ ($\Lambda_{c}^{+}$
	and $\Xi_{c}^{+,0}$), we have
	\begin{equation}
	{\cal B}_{cq_{1}q_{2}}^{\bar{\boldsymbol{3}}}=\frac{1}{\sqrt{2}}(c(q_{1}q_{2})_{S}-c(q_{2}q_{1})_{S}),
	\end{equation}
	where $(q_{1},q_{2})=(u,d),(u,s),(d,s)$ for $\Lambda_{c}^{+}$ and
	$\Xi_{c}^{+,0}$, respectively.
	
	For ${\cal B}_{q_{1}q_{1}q_{2}}$ ($p$, $n$, $\Sigma^{+,-}$ and
	$\Xi^{0,-}$), we have
	\begin{equation}
	{\cal B}_{q_{1}q_{1}q_{2}}=\frac{1}{2\sqrt{3}}\left(-q_{1}(q_{1}q_{2})_{A}-q_{1}(q_{2}q_{1})_{A}+2q_{2}(q_{1}q_{1})_{A}+\sqrt{3}q_{1}(q_{1}q_{2})_{S}-\sqrt{3}q_{1}(q_{2}q_{1})_{S}\right),
	\end{equation}
	where $(q_{1},q_{2})=(u,d),(d,u),(u,s)(d,s),(s,u),(s,d)$ for $p$,
	$n$, $\Sigma^{+,-}$, $\Xi^{0,-}$, respectively.
	
	For $\Sigma^{0}$ and $\Lambda$, we have
	\begin{align}
	\Sigma^{0} & =\frac{1}{2\sqrt{6}}(2s(ud)_{A}+2s(du)_{A}-d(us)_{A}-d(su)_{A}-u(ds)_{A}-u(sd)_{A}\nonumber \\
	& \quad\quad+\sqrt{3}u(ds)_{S}-\sqrt{3}u(sd)_{S}+\sqrt{3}d(us)_{S}-\sqrt{3}d(su)_{S}),\\
	\Lambda & =\frac{1}{2\sqrt{6}}(\sqrt{3}d(us)_{A}+\sqrt{3}d(su)_{A}-\sqrt{3}u(ds)_{A}-\sqrt{3}u(sd)_{A}\nonumber \\
	& \quad\quad+2s(ud)_{S}-2s(du)_{S}+d(us)_{S}-d(su)_{S}-u(ds)_{S}+u(sd)_{S}).
	\end{align}

\end{document}